\documentclass{aa}

\usepackage{graphicx}
\usepackage{txfonts}
\usepackage{longtable}
\usepackage{threeparttablex}
\usepackage{caption} 
\begin{document}

   \title{CAPOS: The bulge Cluster APOgee Survey \\
   VIII.  Final ASPCAP results for all clusters}

   \titlerunning{CAPOS VIII. Final ASPCAP results}
   \author{Doug Geisler \inst{1,2}, 
   Cesar Mu\~noz \inst{2}, Sandro Villanova \inst{3}, Roger E. Cohen \inst{4},  
   Dante Minniti \inst{5,6},
    Antonela Monachesi \inst{2},
Steven R. Majewski \inst{7},
    Andrea Kunder \inst{8}, Beatriz Barbuy \inst{9}, Katia Cunha \inst{10,11}, Verne Smith \inst{12}, 
    Carolina Montecinos
    \inst{2}, Wisthon Haro Moya \inst{2}, Nicolas Barrera \inst{2},
    Matias Bla\~na \inst{13},
}
   
   \authorrunning{Geisler et al.}

 \institute{Departamento de Astronom\'{i}a, Casilla 160-C, Universidad de
  Concepci\'{o}n, Concepci\'{o}n, Chile
\and  Departamento de Astronomía, Facultad de Ciencias, Universidad de La Serena. Av. Raul Bitran 1305, La Serena, Chile 
\and Universidad Andres Bello, Facultad de Ciencias Exactas, Departamento de Ciencias Físicas - Instituto de Astrofisica, Autopista Concepcion-Talcahuano 7100, Talcahuano, Chile
\and Deptartment of Physics and Astronomy, Rutgers the State University of New Jersey, 136 Frelinghuysen Road., Piscataway,
NJ 08854, USA
\and Instituto de Astrofísica, Depto. de Física y Astronomía,  Facultad de Ciencias Exactas, Universidad Andres Bello, Av. Fernandez Concha 700,  Santiago, Chile
\and Vatican Observatory, V00120 Vatican City State, Italy
\and Department of Astronomy, University of Virginia, Charlottesville,
VA 22904-4325, USA
\and Saint Martin's University, 5000 Abbey Way SE, Lacey, WA 98503, USA
\and Universidade de São Paulo, IAG, Rua do Matão 1226, Cidade Universitária,
São Paulo 05508-900, Brazil
\and Steward Observatory, The University of Arizona, 933 North Cherry
Avenue, Tucson, AZ 85721-0065, USA
\and Observatório Nacional, Rua General José Cristino, 77, 20921-400
Sao Cristóvao, Rio de Janeiro, RJ, Brazil
\and National Optical Astronomy Observatory, 950 North Cherry
Avenue, Tucson, AZ 85719, USA
\and Vicerrectoría de Investigación y Postgrado, Universidad de La Serena, La Serena, 170000, Chile
}


 
  \abstract
   {Bulge globular clusters (BGCs)
are exceptional tracers of the formation and chemodynamical evolution of this oldest Galactic component.
However, until now, observational difficulties have prevented us from taking full advantage of these powerful Galactic archeological tools.
}
   {CAPOS, the bulge Cluster APOgee Survey, addresses this key topic by observing a large number of BGCs, most of which have only been poorly studied previously. 
 We aim to obtain accurate mean values for metallicity, [$\alpha$/Fe], and radial velocity, as well as abundances for eleven other elements. 
Here we  present final parameters based on ASPCAP for all 18 CAPOS BGCs.}
   {We use atmospheric parameters, abundances and velocities from ASPCAP
    in DR17.}
   {First, we carry out a stringent  selection of cluster members, finding a total of   303 having spectral SNR>70  and an additional 125 with lower SNR. 
We reinforced the finding  from multiple population studies that stars with high [N/Fe] abundances show
higher [Fe/H] than their lower [N/Fe] counterparts. Mg, Ca and global $\alpha$ abundances show similar trends, while Si is well-behaved.
The [Fe/H] value of these  2nd population stars is corrected to derive the  mean metallicity. 
Mean metallicities are determined to a precision of 0.05 dex, [$\alpha$/Fe] to 0.06 dex, and
radial velocity to 3.4 km/s.
No clusters show
strong evidence for internal metallicity variations , including M22.
Abundances for eleven other elements using only 1st population stars are calculated.  Our values are generally in good agreement with the  literature.
We developed a new chemodynamical GC classification
scheme, synthesizing  several recent  studies. We also compile  up-to-date metallicities. The BGC metallicity distribution is  bimodal, with peaks near [Fe/H] = -0.45 and -1.1, with the metal-poor peak strongly dominant. The entire in situ sample, including Disk and BGCs, shows the same bimodality, while ex situ GCs are  unimodal, with a peak around -1.6. 
Surprisingly, we find  only a small and statistically insignificant difference in the mean [Si/Fe] of in and ex situ GCs.  
 The 4 GCs with the lowest [Si/Fe] values are
all ex situ, relatively young, and 3 belong to Sagittarius, but no other correlations are evident.

}
{}


   \maketitle
   
   \titlerunning{test}
   \authorrunning{test}
%

\section{Introduction}
Understanding galaxy formation and evolution is a key goal of modern astronomy. A critical target is our own Milky Way (MW) Galaxy. To piece together  the physical processes involved and how they interacted requires probing witnesses to the ancient epoch
of MW formation, which still survive today  to bear testimony, and comparing the evidence derived from these witnesses to theoretical simulations. Globular clusters (GCs) fully fulfill the requisite attributes of ideal witnesses.  They are  bright (with some low-luminosity as well), numerous, inhabit all major MW components (bulge, disk and halo), and, most importantly, ancient yet yield very precise and accurate ages. Hence, they are among
our most powerful Galactic archaeology tracers.

In recent years, we have
learned a tremendous amount about the history of the halo from its GCs, which are the most accessible GC system, with little or no extinction, and are thus the classic prototype GC population. However, these observations as well as simulations now show the halo was mostly accreted (Tumlinson 2010, Massari et al. 2019).
Thus, studying the halo GCs tells us a great deal about the accretion history of the MW but not about the origin of its native, in situ population. For this, we must turn to the Galactic bulge (GB) and Galactic disk (GD), which simulations strongly suggest were mainly formed in situ (Gargiulo et al. 2019).  Thus, it is imperative to investigate GB/GD GCs (B/D GCs) to reveal the oldest, in situ Galactic component, referred to hereafter as the main progenitor, and its past. 

Fortunately, both the GB and GD possess significant GC populations, poised to tell us about the most innate nature of our Galaxy.  
Belokurov \& Kravtsov (2024) classify fully 2/3 of Galactic GCs as in situ.
This ratio will only increase, as significant numbers of new GCs continue to be found, mostly in the bulge or disk.  Recently, Bica et al. (2024) list a total of 39 newly identified GCs and candidates, of which only 7 are associated with the halo.
Unfortunately, until recently we have not been able to unleash
the full power of the B/D GCs  to help unravel the history of the main progenitor due to the 
high extinction and severe crowding toward the plane and central regions of the MW. The recent advent of near-IR high spatial resolution imaging as well as high spectral resolution spectroscopy has finally opened the window to investigating Galactic archaeology in depth via B/D GCs.

CAPOS (the bulge Cluster APOgee Survey - Geisler et al. 2021, hereafter G21) is designed to exploit this spectroscopic opportunity. The primary goal of CAPOS is to obtain detailed abundances and kinematics for a large sample of bona fide members in a number of B/D GCs using the unique advantages of the high resolution (R$\sim$22,500) APOGEE-2S instrument (Wilson et al. 20129) attached to the du Pont 2.5m telescope at Las Campanas Observatory (Bowen and Vaughn 1973) as part of the fourth iteration of the Sloan Digital Sky Survey (SDSS-IV; Blanton et al. 2017).

In the first CAPOS paper (G21), we included an overview of
the project and initial results based on the APOGEE Stellar Parameters and Chemical Abundance Pipeline – ASPCAP
(García Pérez et al. 2016) analysis of the CAPOS data for the seven B/D GCs released in DR16 (Ahumada et al. 2020). We deferred treatment of the eighth cluster, NGC 6656 - M22 - due to its likely ex situ (Perez-Villegas et al. 2020 - hereafter PV20) and metal-complex (Norris \& Freeman
1983) nature. There we described in detail cluster and cluster membership selection for all CAPOS fields available at that time, as well as the selection of bulge field stars, K2 stars (Stello et al. 2017), EMBLA stars (Howes et al. 2016) and PIGS stars (Arentsen et al. 2020). These field stars were included to fill fibers that could not be targeted on GCs due to fiber collision limits with
higher priority cluster targets.

In our detailed ASPCAP abundance analysis in G21, we found a trend within each of our clusters for stars with
the highest N abundances (termed originally 2G stars but here referred to as 2P stars, to link them to 2nd generation/population stars in the multiple population [MP] phenomenon) to also show a higher
metallicity than their 1G/1P (1st generation/population) counterparts. We interpret this as due
to a known issue with ASPCAP to overestimate effective temperatures
for 2P stars, with their unique abundance ratios. We used the metallicity of 1P stars to
correct [Fe/H] values for 2P stars, and used these corrected values
together with the uncorrected 1P values to derive the mean [Fe/H]
for our clusters. We also calculated mean abundances for eleven other elements in 1P stars not expected to be affected by MP,
the mean [$\alpha$/Fe] and radial velocity, and explored MP in the seven 
CAPOS B/D GCs investigated.  

DR16 contained APOGEE data taken by June 2019, which only included three CAPOS fields. These three fields include, as already mentioned, seven B/D GCs, NGC 6656, and a candidate GC from Minniti et al. (2017). Subsequent CAPOS observations were obtained in July 2019 and October 2020 for 10 additional B/D GCs and 2 additional Minniti et al. candidates in 5 new CAPOS fields, as well as additional exposures for some of the objects studied in G21.

Here we present results for all 18 CAPOS clusters, including NGC 6656, contained in DR17 (Abdurro'uf et al. 2022), which comprises all APOGEE data obtained for this project. We basically follow the same procedure as in G21, first 
discussing the cluster sample and our observations (Section 2) and membership selection (Section 3).
We then consider the atmospheric parameters, including 
investigating any effect of MP on the metallicity and other elemental abundances (Section 4).
Next, we derive mean  metal and  [$\alpha /Fe]$  abundances and radial velocity (Section 5) and other mean elemental abundances (Section 6) for each cluster, utilizing ASPCAP abundances from DR17, and compare our values to previous studies. Section 7 examines the salient implications of these results, including the nature of our sample, which presents a revised chemo-dynamical classification for all Galactic GCs, a discussion of the BGC metallicity distribution (MD), and a comparison of the MDs and [$\alpha /Fe]$ distributions of in situ vs. ex situ GCs. Finally, Section 8 summarizes our work. An Appendix investigates the reality of the 3 Minniti et al. candidates.
We note that other papers in this series present results derived for some CAPOS GCs using independent photometric atmospheric parameters and an abundance analysis using the BACCHUS code (Masseron et al. 2016), and we include a comparison to these studies.

\section{Cluster sample and observations}

The CAPOS cluster selection was described in detail in G21. Briefly, our targets were presumed bulge GCs (BGCs) from a variety of sources (see details in G21), lying within the innermost $\pm 10^\circ$ in latitude and longitude around the Galactic center, and not planned to be observed with APOGEE-2S as part of the main SDSS-IV survey. Indeed, as shown in Figure 1 of G21, although the SDSS-IV APOGEE survey (Majewski et al. 2017) included a large number of fields in the central  10$^\circ$, only a small number of BGCs were actually observed by the survey.
Our observations were generally obtained before Gaia DR1 and thus 
we limited our definition of BGCs to their three-dimensional spatial location, using
a 3.5 kpc Galactocentric distance maximum limit for the bulge. We ignored cluster metallicity, orbit, and origin, since these latter two were essentially unknown until the advent of Gaia data,
while the metallicity distribution function of bonafide BGCs was, and still is, uncertain, especially the lower limit. We prioritized fields with multiple BGCs.  The last field observed, containing NGC 6717, is a few degrees beyond our nominal 10$^\circ$ limit, and was selected for its extreme right ascension, and in fact was the last bulge field observed by APOGEE-2S as part of SDSS-IV.
Although NGC 6656 is outside of our limiting Galactocentric
radius for BGCs, it is an interesting GC in its own right,
having been the subject of considerable debate as to whether or
not it has an intrinsic metallicity spread (e.g., Norris \& Freeman
1983; Mucciarelli et al. 2015), and was readily observed simultaneously
with the BGC NGC 6642.

Table 1 gives basic positional data for all CAPOS clusters, including
the APOGEE field ID, while Table 2 lists additional basic parameters
from the literature for the clusters.
The final CAPOS sample now includes a total of 18 GCs. 
Note that we have added FSR 1758, which was not included in the same table in G21, as it did not appear in the Harris (1996 - hereafter H10) catalog at the time of our original selection but was discovered subsequently (Barb\'a et al. 2019) and happily falls within the same APOGEE field as two other cataloged BGCs and we were able to target possible members in time for our observations.

We also obtained data for three candidate GCs from Minniti et al. (2017): dubbed Minni 6, 51 and 66. These will be discussed briefly in an Appendix but are not included in our final compilation as none
of these candidates show evidence from our study for being a star cluster. 
We also note that, while not all of our clusters are now considered bonafide BGCs, 14 of the original 16 considered to be BGCs are 
indeed now classified by us (Section 7) as bonafide BGCs, while the other two are  Disk GCs (DGCs). The two extra GCs observed "for free" in the same fields as our main targets are classified as ex situ (FSR 1758) or unknown  as to whether they are in or ex situ (NGC 6656), as we also discuss below.

Finally, although some of our clusters,
and indeed some of our actual cluster members, were also observed by other SDSS-IV programs with APOGEE, we opt to present here only data obtained by the CAPOS survey.
We also do not discuss results from any of the field stars targeted in CAPOS, leaving these for subsequent papers.

\section{Cluster membership selection}

In G21, final membership selection was made post-observation using the APOGEE DR16 catalog, where we utilized ASPCAP radial velocities (RVs) and metallicities and Gaia DR2 proper motions (PMs) as membership criteria. 
For the abundance analysis, we eliminated potential members with signal-to-noise ratio (SNR)$<$70 for consistency with other GC papers based on APOGEE (e.g., Meszaros et al. 2020). We have optimized and improved our selection procedure in this study. For this purpose, we have employed the APOGEE DR17 catalog, focusing on the parameters provided by the ASPCAP pipeline. This release includes   more fields, more clusters, more stars, and, in some cases, a better SNR for some previously selected stars than available in G21. Additionally, we have incorporated information provided by Gaia DR3 regarding PM, along with the study conducted by  Vasiliev \& Baumgardt (2021) and available in  the dedicated website of the authors,\footnote{https://people.smp.uq.edu.au/HolgerBaumgardt/globular/} which primarily uses Gaia DR3 data to characterize the PM of most of the Galactic GCs. This approach substantially improves our new membership selection.

APOGEE DR17 provides precise RVs, accurate to typically $\sim$0.05 km $s^{-1}$, and metallicities determined by ASPCAP with a precision better than 0.10 dex, for stars with SNR$\geq 70$, which provides key membership information. Additionally, Gaia DR3 has provided astrometric data for practically all stars observed by APOGEE. Note that, although the CAPOS clusters are generally quite reddened, the stars observed by APOGEE were limited to the brightest few magnitudes of the RGB, ensuring that Gaia data were available for all stars in our sample. Table 1  lists each selected GC in this study with its corresponding APOGEE field, which in most cases contains more than one CAPOS GC.

\begin{figure*}
\centering
\includegraphics[width=15cm]{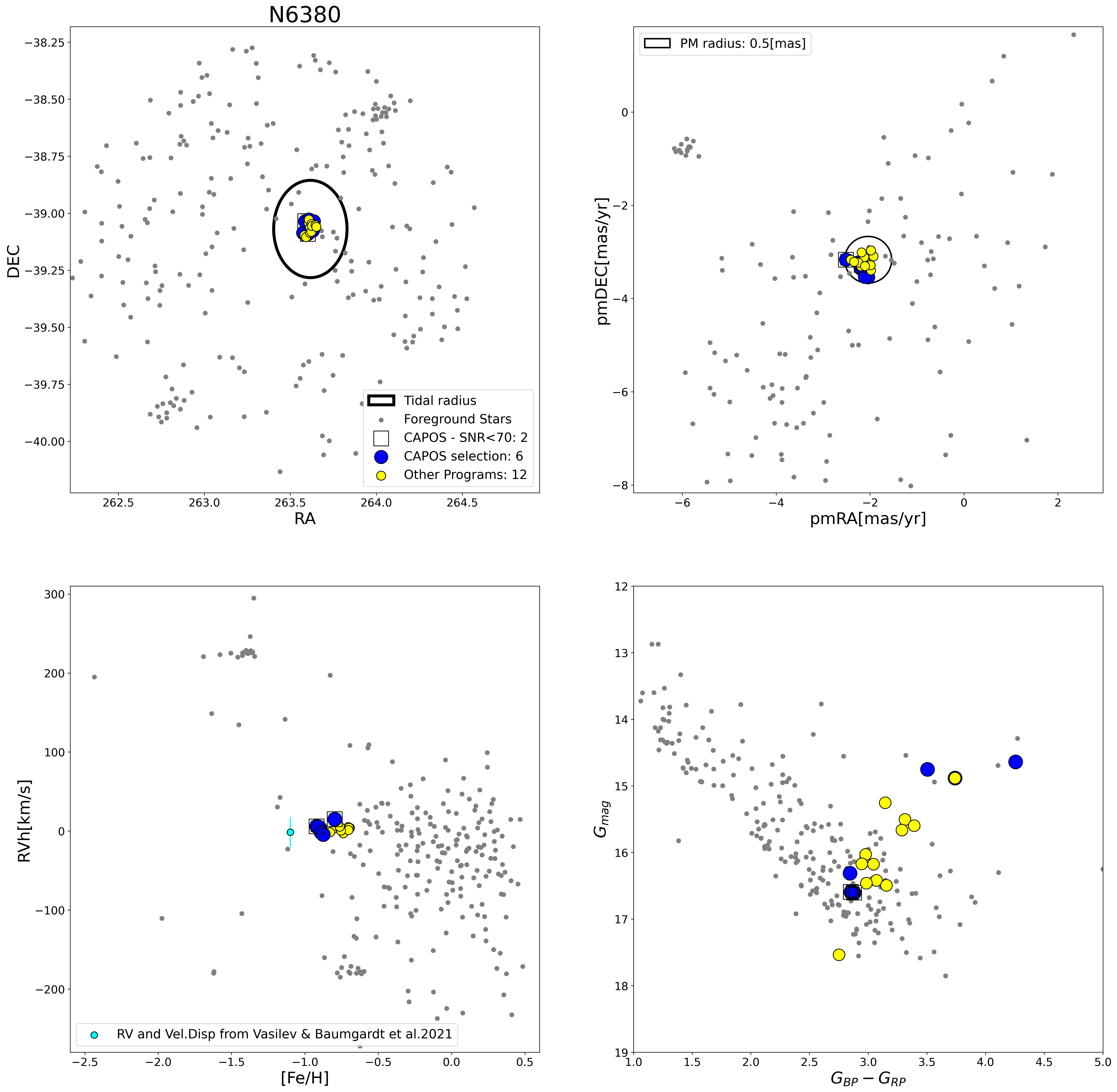}
      \caption{Membership determination in NGC 6380, including location with respect to tidal radius (upper left), PM space (upper right), metallicity:RV space (lower left) and CMD (lower right). Shown are CAPOS final members in blue, with boxes around those with SNR$<$70, non-CAPOS members in yellow, and other stars observed by APOGEE in grey. The Vasiliev \& Baumgardt (2021) mean RV and error are shown in cyan. The clustering seen in the grey points are due to two other GCs observed in the same CAPOS field. }
    \label{N6380members}
\end{figure*}

Below, we detail step by step the membership selection process. Figure \ref{N6380members}
 graphically illustrates our criteria and procedure for one of our sample, NGC 6380. Note that this cluster is not very offset
from the field in terms of RV or PM (in contrast to the other two clusters in this APOGEE field), but
our procedure is still able to pick out the cluster stars with high probability.

First, we selected only stars within the tidal radius, for which we used the value defined by Vasiliev \& Baumgardt  (2021).

Second, we have used the PM from Gaia DR3 for each star along with the determination of the mean PM for each  GC determined by Vasiliev \& Baumgardt  (2021). We used a maximum radius of 0.5 mas $yr^{-1}$  from the PM centroid for  almost all CAPOS GCs  to maintain homogeneity. Our analysis shows that this radius is optimal for maximizing members and minimizing contamination from field stars. However, for the lone case of NGC 6656, we extended this radius to 1.0 mas yr$^{-1}$ for two reasons. First, this cluster presents around 400 potential members in APOGEE DR17, with many likely members from the other criteria lying between 0.5 - 1.0 mas yr$^{-1}$. Second, this GC is relatively isolated from field stars in the PM diagram. Thus, contamination from field stars is minimal, even out to 1 mas $yr^{-1}$.

For a third constraint, we used the ASPCAP RVs. These were compared with the mean and dispersion determined by Vasiliev \& Baumgardt  (2021). In Figure \ref{N6380members}, in the bottom left panel, their mean is shown in cyan, along with its corresponding dispersion represented by the error bars. For our selection, we have considered stars within $2\sigma$ from the mean.

The fourth criterion was metallicity. In general, the average metallicity of the CAPOS GCs has been previously derived using a variety of methods. H10 lists these metallicities in his catalog. In addition to this, considering that GCs generally do not exhibit a large dispersion in metallicity, and taking into account that ASPCAP provides an uncertainty of $\sim$0.10 dex, we have considered stars with a metallicity given by ASPCAP $<3\sigma$ from the cluster mean given by Vasiliev \& Baumgardt  (2021) as members. Additionally, we verified that all our targets are located in the main branches (RGB or AGB) corresponding to the CMD of each cluster.



Finally, we have used all members irrespective of their SNR to calculate the mean RVs of each GC, using a mean weighted by their SNR (see Table 4). It is important to note, however, that the star with the lowest SNR in the entire sample has a value of 29, and several authors have demonstrated that even stars with lower APOGEE SNRs can still provide excellent RV values (Munoz et al. 2023, Nidever et al. 2020). However, we have separated the stars with SNR>70 as these are the only ones used for the determination of chemical abundances, following the recommendation of previous studies such as Meszaros et al. (2020) and in keeping with G21.
In Table 3, we enumerate the total number of stars selected for each GC  in the column labeled `all CAPOS stars', as well as the number with SNR$>$ 70, and the mean metallicity we derive. 

This extensive, multidimensional culling procedure ensures very high quality membership selection for our final surviving candidates.
 The final CAPOS cluster members are available at the CDS 
via anonymous ftp  to cdsarc.cds.unistra.fr (130.79.128.5) 
or via https://cdsarc.cds.unistra.fr/viz-bin/cat/J/A+A/Vol/Num.

\subsection{Comparison to Schiavon et al. (2024) VAC}

Additionally, we have compared our selection with that made by Schiavon et al. (2024 - hereafter S24), who have also used the same DR17 ASPCAP catalog derived from the APOGEE survey. Note that their study encompasses all Galactic GCs observed by APOGEE, unlike this study, which focuses only on the CAPOS GCs. In addition, our selection is based only on stars observed for the "CAPOS" program,  while S24 uses all available stars without discriminating between programs; i.e., for a few of our GCs, observations were obtained with APOGEE independently of CAPOS. 
 Note that S24 specifically designed their study to "generously consider every star with a reasonable probability of
belonging to a given GC, providing elements to enable the catalogue users to make their own informed sample selections. In short, catalogue completeness is prioritized over purity."
In contrast, we have prioritized purity over completeness, so one would expect our sample to in general have fewer candidate members than S24 and that some of their candidates are likely non-members. In fact, the VAC would have been an excellent starting point for our membership selection, but  we completed our selection before S24 was published.

The comparison is shown in Table 3. In general, we find good agreement with the results of S24. However,  as expected,
we have found some differences between stars selected in the two studies for several clusters. The main differences are due to stars selected by S24 that lie outside the tidal radius that we have used in this study, or fall  beyond the 0.5  mas yr$^{-1}$ PM radius defined by us. In these cases, selection clearly depends on the criteria adopted by each author. In this study, we have used the tidal radius and  mean PM values defined by  Vasiliev \& Baumgardt  (2021). In addition, we find that some stars selected by S24 present a metallicity or RV determined by ASPCAP that are in   substantial disagreement with  our GC mean.  Note that S24 did not include a metallicity criterion for all of their candidates in order to to avoid missing members for which ASPCAP could not find a metallicity solution or those with potentially large errors in metallicity, and to also allow the possibility of candidates with a real metallicity spread.

\begin{figure*}
\centering
\includegraphics[width=15cm]{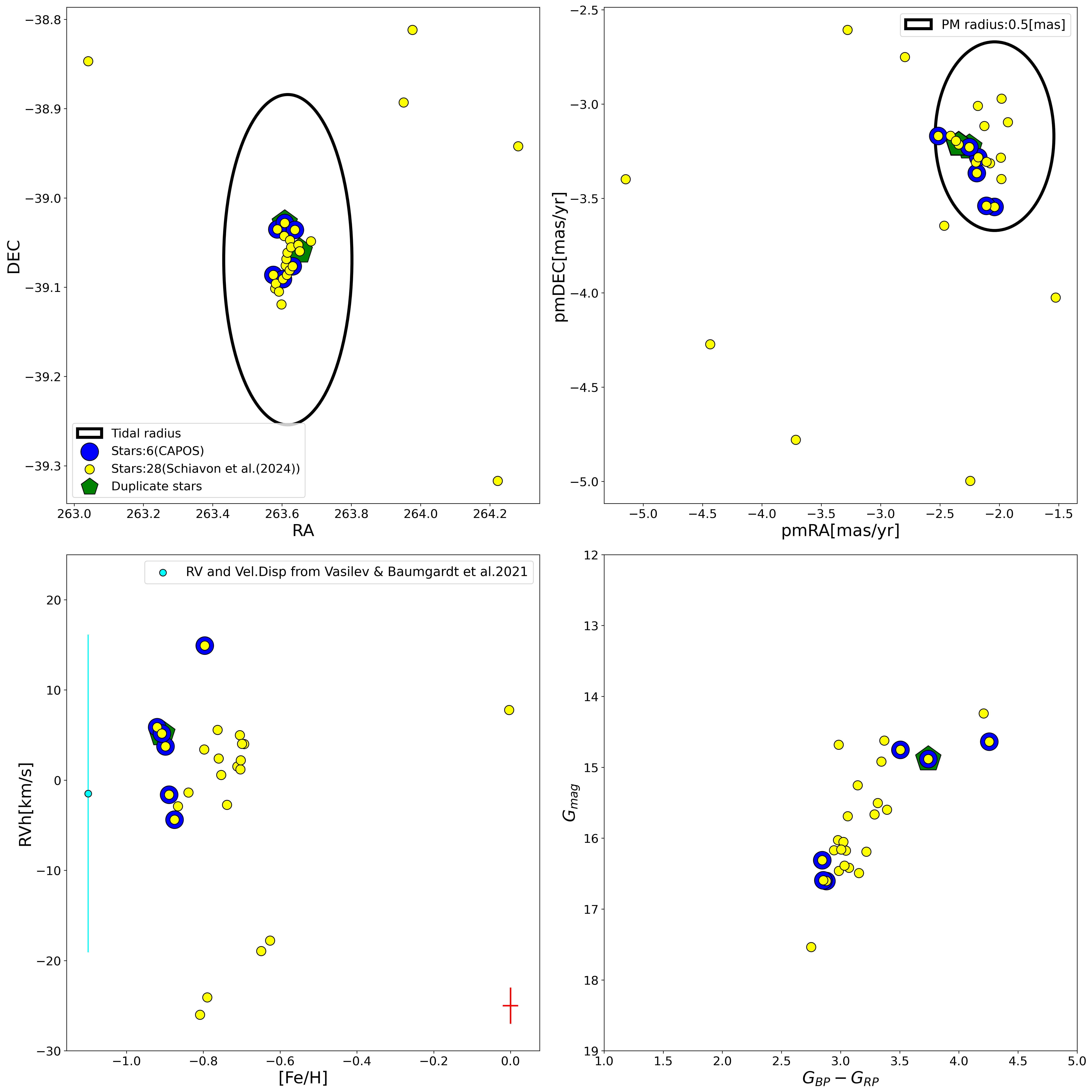}
      \caption{Comparing our membership determination in NGC 6380 vs. that of S24. Diagrams and point descriptions as in Figure \ref{N6380members}, but with yellow points indicating members selected by S24, and green pentagons showing duplications in their selection. The typical error in PM is about 0.01 mas $yr^{-1}$ for both pmRa and pmDec, and in the RV vs metallicity plot,  typical errors are shown in red. S24 have included a number of stars which do not satisfy our more stringent membership requirements.}
    \label{N6380Schiavon}
\end{figure*}

In Figure \ref{N6380Schiavon}, we present an example of the comparison for the cluster NGC 6380, where we show both selections. We note several clear differences in the selection. First, four stars chosen by S24 are located outside the tidal radius of 14.1 arcmin adopted here. Also, eight stars lie substantially more than 0.5 mas yr$^{-1}$ from the center of the GC in the PM diagram. Finally, in the lower left panel of Figure \ref{N6380Schiavon}, where the RV vs metallicity plot is presented, we note that several stars have a RV and metallicity exceeding the limit defined by us with respect to the mean.  This clearly leads to a difference in mean metallicities between the two studies; in this case, the mean metallicity derived by S24 is -0.78 dex, contrasting with our significantly more metal-poor value of -0.90 dex (see Table 3).
Again, such differences are to be expected, considering that S24 aims to include the largest possible number of cluster members with a reasonable probability.


Finally, we have found that some stars reported as members in S24 are duplicated or even triplicated. This situation was acknowledged by S24 and is mainly due to the fact that there are multiple entries for stars located in overlapping fields and observed as part of different programs.
The CAPOS GCs with repeated members in the S24 catalogue are HP1, Terzan 9, NGC 6380, and NGC 6656. It is quite likely that such duplication, as well as the inclusion of low membership probability stars described above, also occurs in other non-CAPOS GCs.  However, field star contamination increases significantly toward the bulge, which is precisely where this study focuses its analysis, and contamination  is certainly far less problematic at high Galactic latitude, where the remaining APOGEE VAC sample is mostly located.

\begin{table*}
\centering
\begin{threeparttable}[b]
\label {basicparamsall}
\caption{Basic positional data for all CAPOS clusters.}

\centering
\small
\begin{tabular}{ c c c c c c }
 
\hline
Cluster ID & $\alpha$ (J2000.0) & $\delta$ (J2000) & l($^\circ $) & b($^\circ $) & APOGEE  \\
 &   $hh:mm:ss$ & $^\circ $~:~ $ ' $~: ~ $''$  &  & & Field  \\

\hline
NGC 6273 & 17 02 37.8 & -26 16 04.7  & 356.87  &  9.38& 357+09-C \\
NGC 6293  &  17 10 10.2 & -26 34 55.5  & 357.62  &  7.83 & 357+09-C \\
NGC 6304 & 17 14 32.3 &  -29 27 43.3  & 355.83  &  5.38 & 356+06-C \\
NGC 6316 & 17 16 37.3 &  -28 08 24.4  & 357.18  &  5.76 & 356+06-C \\
Terzan 2 & 17 27 33.1 & $-$30 48 08.4 & 356.32  &  2.30 & 357+02-C \\
Terzan 4 &  17 30 39.0 & $-$31 35 43.9  & 
356.02  &  1.31 & 357+02-C \\
HP1	&  17 31 05.2 & $-$29 58 54 & 
357.44  &  2.12 & 357+02-C \\
FSR 1758 & 17 31 12.0 &  -39 48 30  &   349.22 &  -3.29 & 350-03-C \\
NGC 6380 & 17 34 28.0 &  -39 04 09  &   350.18 &  -3.42 & 350-03-C \\
Ton 2 & 17 36 10.5  & -38 33 12     & 350.80  & -3.42 & 350-03-C \\
Terzan 9 &  18 01 38.8 & $-$26 50 23 & 
3.61  & -1.99 & 003-03-C \\
Djorg 2	& 18 01 49.1 & $-$27 49 33 & 
2.77  & -2.50 & 003-03-C \\
NGC 6540	&  18 06 08.6 & $-$27 45 55 & 3.29  & -3.31 & 003-03-C \\
NGC 6558 & 18 10 17.6 &  -31 45 50.0  &   0.20  & -6.02 & 000-06-C \\
NGC 6569 & 18 13 38.8 &  -31 49 36.8  &   0.48  & -6.68 & 000-06-C \\
NGC 6642	& 18 31 54.1 & $-$23 28 30.7 & 9.81 & -6.44 & 010-07-C \\
NGC 6656	&  18 36 23.9 & $-$23 54 17.1 & 9.89 &  -7.55 & 010-07-C \\
NGC 6717  & 18 55 06.0 &  -22 42 05.3  &  12.88 & -10.90 & 013-11-C \\

\hline
\end{tabular}

Equatorial coordinates, l and b are from H10 
(except for FSR 1758, whose coordinates are taken from Cantat-Gaudin et al. 2018). 

\end{threeparttable}
\end{table*}


\begin{table*}
\centering
\begin{threeparttable}[b]
\label {basicparams}
\caption{Basic parameters from the literature for all CAPOS clusters.}

\centering
\small
\begin{tabular}{ c c c c c c c}
 
\hline
Cluster ID & [Fe/H] & E(B-V) & $V_r$ & $\mu _{\alpha} cos\delta$ & $\mu _{\delta}$ &Mass \\
 &    & & km $s^{-1}$  & mas $yr^{-1}$ & mas $yr^{-1}$ & M\(\odot\)\\

\hline
NGC 6273 &   -1.74 & 0.38 & 145.54$\pm$0.59 & -3.249$\pm$0.026 &  1.660$\pm$0.025 &7.19 $\pm$0.37 $\times$ 10$^{5}$  \\
NGC 6293 &   -1.99 & 0.36 & -143.66$\pm$0.39 & 0.870$\pm$0.028 &  -4.326$\pm$0.028&1.42 $\pm$ 0.12 $\times$ 10$^{5}$ \\
NGC 6304 &   -0.45 & 0.54 & -108.62$\pm$0.39 & -4.070$\pm$0.029 &  -1.088$\pm$0.028&1.03 $\pm$ 0.04 $\times$ 10$^{5}$\\
NGC 6316 &   -0.45 & 0.54 & 99.81$\pm$0.82 & -4.969$\pm$0.031 &  -4.592$\pm$0.030&3.47 $\pm$  0.44 $\times$ 10$^{5}$\\
Terzan 2 &   -0.69 & 1.87 & 128.96$\pm$1.18 & -2.170$\pm$0.041 &  -6.263$\pm$0.038&8.05 $\pm$ 2.31 $\times$ 10$^{4}$\\
Terzan 4 &  -1.41 & 2.00 & -39.93$\pm$3.76 & -5.462$\pm$0.060 &  -3.711$\pm$0.048&1.81 $\pm$ 0.53 $\times$  10$^{5}$\\
HP1	&  -1.00  & 1.12 & 40.61$\pm$1.29 & 2.523$\pm$0.039 &  -10.093$\pm$0.037&1.37 $\pm$ 0.19 $\times$ 10$^{5}$\\
FSR 1758 &   -1.50 & 0.90 & 226.8$\pm$1.6 & -2.881$\pm$0.026 &  2.519$\pm$0.025&4.91 $\pm$  0.51 $\times$ 10$^{5}$\\
NGC 6380 &   -0.75 & 1.17 & -6.54$\pm$1.48 & -2.183$\pm$0.031 &  -3.233$\pm$0.030&3.41 $\pm$0.05 $\times$  10$^{5}$\\
Ton 2 &   -0.70 & 1.24 & -184.72$\pm$1.12 & -5.904$\pm$0.031 &  -0.755$\pm$0.029&4.31 $\pm$ 0.91 $\times$ 10$^{4}$\\
Terzan 9 &   -1.05  & 1.76 & 29.31$\pm$2.96 & -2.121$\pm$0.052 &  -7.763$\pm$0.049&1.37 $\pm$ 0.16 $\times$ 10$^{5}$	\\
Djorg 2	&  -0.65 & 0.94 & -148.05$\pm$1.38 & 0.662$\pm$0.042 &  -2.983$\pm$0.037&1.34 $\pm$ 0.24 $\times$ 10$^{5}$	\\
NGC 6540	&   -1.35 & 0.66 & -17.98$\pm$0.84 & -3.702$\pm$0.032 &  -2.791$\pm$0.032& 5.62 $\pm$ 0.37 $\times$ 10$^{4}$\\
NGC 6558 &   -1.32 & 0.44 & -194.69$\pm$0.81 & -1.720$\pm$0.036 &  -4.144$\pm$0.034&3.13 $\pm$ 0.10 $\times$ 10$^{4}$\\
NGC 6569 &   -0.76 & 0.53 & -49.83$\pm$0.50 & -4.125$\pm$0.028 &  -7.354$\pm$0.028&2.29 $\pm$ 0.21 $\times$ 10$^{5}$\\
NGC 6642	&  -1.26 & 0.40 & -33.23$\pm$1.13 & -0.173$\pm$0.030 &  -3.892$\pm$0.030&3.95 $\pm$ 0.13 $\times$ 10$^{4}$\\
NGC 6656 &   -1.70 & 0.34 & -147.76$\pm$0.30 & 9.851$\pm$0.023 &  -5.617$\pm$0.023&4.70 $\pm$0.06 $\times$ 10$^{5}$\\
NGC 6717 &   -1.26 & 0.22 & 32.45$\pm$1.44 & -3.125$\pm$0.027 &  -5.008$\pm$0.027&2.63 $\pm$ 0.19 $\times$ 10$^{4}$\\

\hline
\end{tabular}

[Fe/H] and E(B-V) are from H10 
(except for FSR 1758, whose values are taken from Barb\'a et al.  2019). $V_r$, proper motions and mass are from Baumgardt et al. (2019) and Vasiliev \& Baumgardt (2021), respectively
(except for FSR 1758, whose $V_r$ is from Villanova et al. 2019).

\end{threeparttable}
\end{table*}

\begin{table*}
\centering
\begin{threeparttable}[b]
\caption{Comparison with Schiavon et al. (2024) sample and mean metallicity.}
\label{comp_capos_schiavon}

\begin{tabular}{ccccccc}
\hline 
\multicolumn{1}{l|}{}         & \multicolumn{3}{c|}{CAPOS}    & \multicolumn{2}{c|}{Schiavon}    & \multicolumn{1}{c}{Stars in Common}   \\

\multicolumn{1}{c|}{Cluster ID}          & All CAPOS stars & SNR > 70& \multicolumn{1}{l|}{[Fe/H]} &Nro. stars & \multicolumn{1}{c|}{[Fe/H]} & CAPOS/Schiavon  \\ 
\hline
N6273   & 76  & 62 & -1.75$\pm$0.11 & 81 & -1.71 & 76 \\
N6293   & 20  & 13 & -2.12$\pm$0.08 & 20 & -2.09 & 20 \\
N6304   & 29  & 12 & -0.49$\pm$0.06 & 34 & -0.48 & 29 \\
N6316   & 17  & 6  & -0.83$\pm$0.05 & 24 & -0.77 & 17 \\
Ter 2   & 4   & 4*  & -0.88$\pm$0.02 & 5  & -0.86 & 4 \\
Ter 4   & 3   & 3  & -1.41$\pm$0.04 & 3  & -1.38 & 3 \\
HP1     & 10  & 10 & -1.23$\pm$0.07 & 17 & -1.21 & 10 \\
FSR1758 & 15  & 9  & -1.48$\pm$0.08 & 15 & -1.42 & 15 \\
N6380   & 6   & 4  & -0.90$\pm$0.02 & 28 & -0.78 & 6 \\

TON2    & 12  & 6  & -0.73$\pm$0.03 & 11 & -0.74 & 10 \\
Ter 9   & 9   & 9  & -1.42$\pm$0.04 & 23 & -1.36 & 9 \\
Djorg2  & 6   & 6  & -1.14$\pm$0.04 & 10 & -1.07 & 6 \\
N6540   & 4   & 4  & -1.09$\pm$0.06 & 6  & -1.02 & 4 \\
N6558   & 5   & 4  & -1.15$\pm$0.03 & 6  & -0.99 & 5 \\
N6569   &  12   & 9  & -1.04$\pm$0.05 & 14 & -0.92 & 12 \\

N6642   & 11  & 10 & -1.11$\pm$0.04 & 12 & -1.09 & 11 \\
N6656     & 185  & 130 & -1.75$\pm$0.10 & 412 & -1.70 & 185 \\

N6717   & 4   & 2  & -1.17$\pm$0.05 & 5  & -1.12 & 4 \\
\hline
\end{tabular}
*In this case, only 3 stars were used  because ASPCAP does not provide abundance measurements for one of the members.
\end{threeparttable}

\end{table*}

\section{Atmospheric parameters}
The ASPCAP pipeline derives stellar parameters and metallicities from a global fit to the entire spectrum,
and then detailed abundances for more than 20 elements by fitting
the spectral lines to models using these atmospheric parameters.
We note that here we use the ASPCAP-calibrated, spectroscopic atmospheric parameters and abundances.

In G21, while investigating the reliability of these parameters, we found a significant positive gradient of increasing metallicity with
temperature within a cluster, of similar magnitude for most clusters, although data were sparse. We also found a clear trend within each of our clusters for stars with the highest
[N/Fe] abundances to also show a higher metallicity than their cluster
counterparts with lower [N/Fe].
This behavior is consistent with the Jönsson et al. (2018) finding that ASPCAP overestimates (with respect to the optical studies taken as
reference) effective temperatures as well as surface gravities for
so-called 2P stars, i.e., stars with abundance
patterns typical of stars believed to have been born from gas polluted
by evolved 1P stars to form another population(s), leaving a present-day cluster displaying MP, which are found in essentially all GCs above a minimum mass or mass density threshold
(Caretta et al. 2009, Huang et al. 2024).  We note that none of the scenarios so far proposed to explain MPs is able to fully account for the wide variety of observational evidence (Bastian \& Lardo 2018).

We reinvestigated this behaviour with the much larger full CAPOS dataset now available. Note that G21 had a total of only 40 cluster members in 7 clusters with available spectra having SNR>70, for a mean of 
5.7 members per cluster, while the current sample includes a total of 
 303 members in 18 clusters with SNR>70, for a mean of  16.8 members per cluster. Thus, the current sample is >7 times larger in total with almost 3 times as many stars per cluster on average, making the full sample much more robust, although we should add that almost 2/3 of the stars in the full sample now come from only 2 clusters - NGC 6273 and NGC 6656. 

\begin{figure*}
\centering
   \includegraphics[width=17cm]   {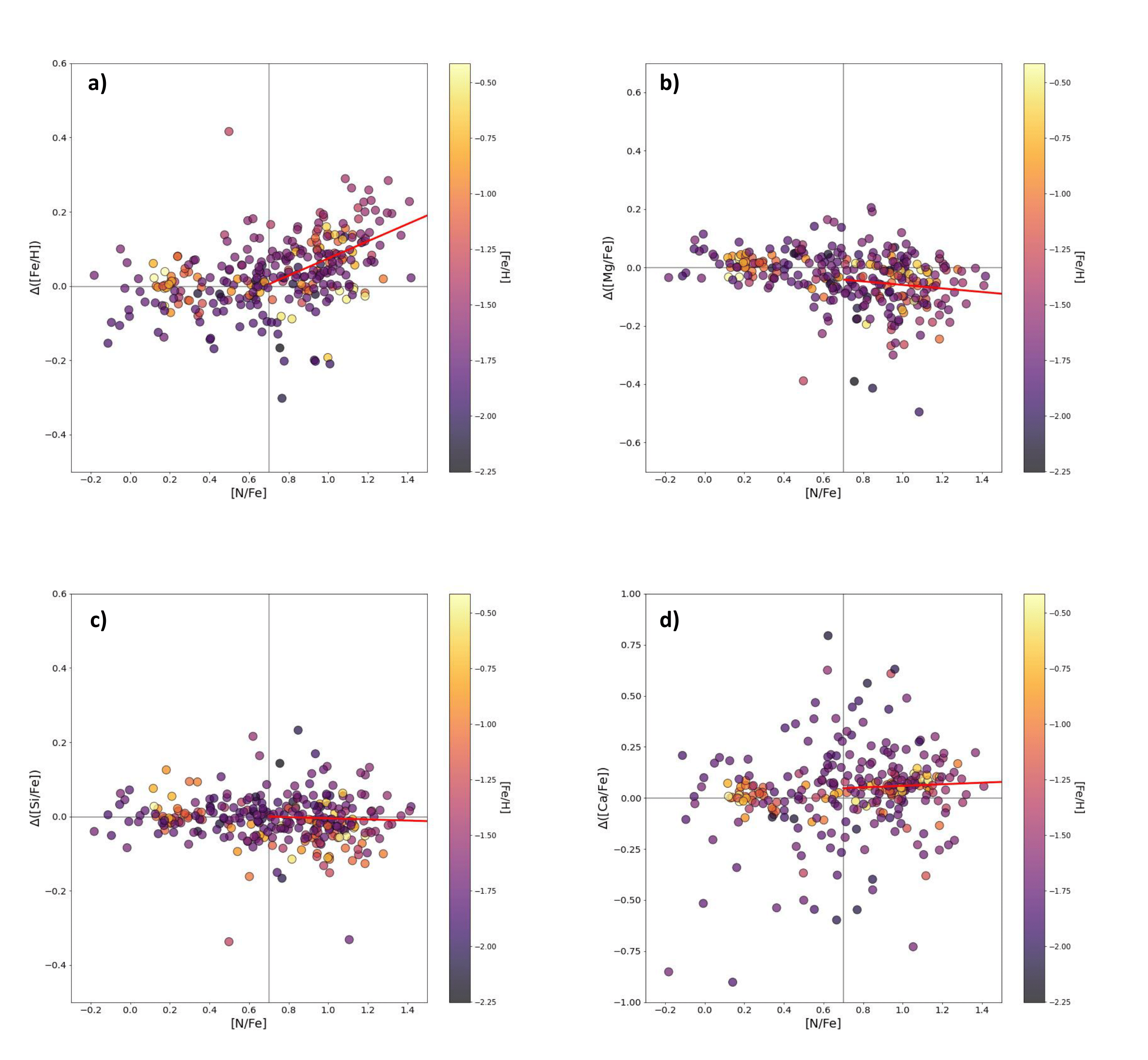}
      \caption{a). $\Delta([Fe/H])$ (difference in [Fe/H] from the mean based on only 1P stars = those with ([N/Fe]$<$0.7) as a function of [N/Fe], 
       A clear trend is found for 2P (higher [N/Fe]) stars to have a higher [Fe/H] than 1P stars in the same cluster, with the difference in metallicity increasing with increasing [N/Fe] for 2P stars. The best-fit line is shown for 2P stars.
      b). As in a) but for [Mg/Fe].
      c). As in a) but for [Si/Fe].
      d). As in a) but for [Ca/Fe]. The stars are color-coded according to metallicity in each panel.
       }
    \label{FevsN}
\end{figure*}

We followed the same procedure as in G21, assigning stars based on the [N/Fe] ratio as either 1P ([N/Fe]<0.7) or 2P
([N/Fe]>0.7). The division at [N/Fe]=0.7, as chosen in G21, again appears very reasonable
(also see Meszaros et al. 2020).  Other studies have suggested a more appropriate value is [N/Fe]=0.5 (e.g. Fernandez-Trincado et al. 2022, S24).  We find that using 0.7 provides a much more statistically significant correlation between $\Delta ([Fe/H]) = [Fe/H]_* - <[Fe/H]>_{1P}$ and [N/Fe]
than using 0.5.
Note that our sample is much larger and covers almost the full metallicity range of GCs  compared to previous studies.
We derived the cluster mean [Fe/H] for 1P stars only, $<[Fe/H]>_{1P}$,  and then plot the difference $\Delta ([Fe/H])$ for all stars in the cluster as a function of [N/Fe]. Figure   \ref{FevsN}
shows our result, which is very similar to Figure 5 of G21 but allows us to discern the trend much more clearly due to the increased sample. In G21, we simply divided all stars into 1P or 2P and found a very consistent mean difference of
$+0.06 \pm 0.01$ dex for all 2P stars compared to 1P stars in the same cluster. We then applied a correction of -0.06 dex to the [Fe/H] value of each 2P star and then used these corrected values together with the 1P values to derive our final mean cluster metallicity: <[Fe/H]>, which was given in Table 3 of G21. 

With our much larger sample, it is now clear that there is not simply a constant offset between 1P and 2P stars but that   
$\Delta([Fe/H])$
increases roughly linearly with [N/Fe] above 0.7. As our GCs now cover a metallicity range >1.6 dex, much larger than in G21, we also investigate any possible metallicity effects in this trend. 
All of our clusters exhibit similar behavior within the errors except our most metal-rich GC - NGC 6304. At [Fe/H]$\sim -0.5$,
it is about 0.25 dex more metal-rich than the next most metal-rich GC (see Table 4) and it is the only GC which shows no significant difference between the metallicities of 1P and 2P stars. 
To derive the final mean cluster metallicity <[Fe/H]>, we simply took the unweighted mean of all stars in NGC 6304, while for the other clusters, we corrected each 2P star by the least-squares line shown in Fig.\ref{FevsN},
based on its [N/Fe] value, and then averaged this corrected value for each 2P star with the (uncorrected) values for all 1P stars. The equation for this line is $\Delta ([Fe/H])$ = 0.22[N/Fe]-0.14 for [N/Fe]$>$0.7.
This mean value is given in Table 4 along with its standard error. Of course, the above assumes that ASPCAP correctly measures [N/Fe]. Further tests (e.g., Montecinos et al. in preparation) comparing ASPCAP to BACCHUS results for 1P and 2P stars suggests that this is indeed a reasonable assumption.

We thus reinforce and clarify the evidence for this behavior found in G21 and the implication that ASPCAP is not optimum for deriving atmospheric parameters for 2P stars, at least not for typical GCs more metal-poor than -0.5, and therefore also not optimum for deriving their detailed abundances, at least Fe. 
 We also investigated what could be responsible for this behaviour.  As part of our BACCHUS analysis for CAPOS GCs, we have generated independent atmospheric parameters for all stars from photometry and isochrones (e.g. Barrera et al. 2025). We have compared the ASPCAP spectroscopic parameters with these photometric parameters for a sample of our stars and find that the offset between them is always larger for 2P stars than 1P stars in both effective temperature and surface gravity, by roughly 100K and 0.2 dex, respectively, with the actual value metallicity-dependent (Montecinos et al. in prep.). This is in accord with the findings of Jonsson et al. (2018). And since ASPCAP overestimates $T_{eff}$ for 2P
stars, it will also overestimate [Fe/H] since higher $T_{eff}$ generally means fainter spectral lines and a higher metallicity is required to fit a given line (G21).

What about other elements?
It is expected, given the
ASPCAP methodology, that there will be systematic deviations
for other elements in 2P stars as well, if the culprit is indeed incorrect atmospheric parameters for these stars. Given the importance of the $\alpha$ elements, especially Mg, Si and Ca within APOGEE, we decided  also  to investigate the behavior of these elements with regard to the effect on ASPCAP abundances of their MP nature and carried out an analogous analysis on these elements. Figure
\ref{FevsN}
also shows  $\Delta ([Mg/Fe])$, $\Delta ([Si/Fe])$ and $\Delta ([Ca/Fe])$
as a function of [N/Fe]. Although the scatter for 2P stars exceeds that of 1P stars, [Si/Fe] shows only a very small systematic deviation for 2P stars ($\leq 0.03$ dex) and only for the most extreme 2P stars.  However, Mg and especially Ca are more problematic.
2P stars show a larger scatter compared to 1P stars than for Si, and they also show a mean linear trend with [N/Fe] that becomes quite large ($\sim 0.1$ dex) for the more extreme 2P stars. Moreover, these lines do not pass thru  $\Delta ([Mg/Fe])$ or 
$\Delta ([Ca/Fe])$ =
0 at the limit ([N/Fe]=0.7) for 2P stars. Ca shows  more scatter for all stars, especially in the clusters more metal-poor than -1.5, and with more metal-rich clusters defining a substantial increase in $\Delta ([Ca/Fe])$ for 2P stars, exceeding 0.1 dex for the most extreme 2P stars.
Finally, we also investigated the global $\alpha$ parameter. 
Its behavior is similar to that of Mg. We conclude that
we can safely use the ASPCAP [Si/Fe] values for all stars but will limit our Mg, Ca and $\alpha$ abundances, as well as abundances for all other elements except Si, to 1P stars only. Of course, [Si/Fe] may indeed be (slightly) enhanced in 2P stars by the MP phenomenon but our results demonstrate that any potential enhancement is virtually negligible, while other elements like Mg are indeed expected to be more affected by MP. In fact, without further analysis one cannot distinguish whether the cause of the distinct abundances for 1P and 2P elements like Mg is due to MP and/or ASPCAP limitations.

Therefore, we will only use 2P stars in this paper to help derive the mean cluster metallicity (using the correction procedure outlined above), the mean [Si/Fe] abundance (without correction),  and of course the mean radial velocity  (using all stars), and only report on mean abundances for other elements for 1P stars. 
As described in detail in G21,
we further restrict ourselves to elements we deem to have well-determined values for 1P stars in ASPCAP,
including C, N, O, Na, Mg, Al, Si, K, Ca, Fe, Mn, and Ni. We do not include Ce, as Hayes et al. (2022) have shown that the ASPCAP Ce abundances are unreliable.
We defer derivation of abundances for 2P stars to other papers using boutique analyses, e.g., BACCHUS,  to obtain more reliable abundances.

\section{Fundamental cluster parameters: Mean metallicities,
[$\alpha$/Fe] abundances, and radial velocities}

The mean metallicity, designated [Fe/H], of a GC is the primary
parameter detailing its chemical composition. The next
most salient composition indicator is the abundance of the $\alpha$ elements relative to Fe,
designated as [$\alpha$/Fe]. Finally, RVs and their dispersions provide crucial information
regarding membership, internal kinematics and the cluster’s
orbit and therefore origin. Despite the critical importance of these parameters,
among the most fundamental for our understanding of a cluster’s
formation and subsequent chemical and dynamical evolution,
the current state of our knowledge of these parameters for
B/D GCs is woefully inadequate. This is particularly true for most of our current sample.
Note that only one of our sample, NGC 6656, was included in the comprehensive APOGEE study of southern clusters in
SDSS-IV by Meszaros et al. (2020), and indeed this is almost certainly NOT a BGC.
We note the work of Schiavon et al. (2017a) who investigated all 5 bulge GCs available in DR12 that were missing from previous APOGEE analyses.


CAPOS was devised to address this deficiency for
as many BGCs as possible, taking advantage of the powerful
APOGEE instrument (Wilson et al. 2019), which was designed to deliver high precision
RVs and abundances for a large number of elements,
including Fe and all of the species considered $\alpha$ elements. In this
section we present and discuss the ASPCAP results for mean [Fe/H], [$\alpha$/Fe], and
RV for the CAPOS clusters.


Table 4 gives our mean values and their standard deviations. Note that the RV was derived from the mean of all CAPOS members  (number given in column 2 of Table 3),  irrespective of SNR, while the abundances were derived only for members with SNR $>$70 (column  3 in Table 3), but with the metallicities of 2P stars ( = number given in column  3 in Table 3  minus number in column 5 in Table 4) corrected via the procedure outlined in Section 4.

\subsection{Metallicity}

G21 emphasized that previous metallicity information for
our clusters, and indeed most B/D GCs, came from a hodgepodge
of sources with a large range of precision and accuracy, but
mostly of relatively poor reliability and/or not involving near-IR capabilities to help mitigate the extinction problem. Although all of
our clusters have been investigated before, the metallicity estimates are generally
based on relatively low-quality indices. Needless to say, such studies are very inhomogeneous. CAPOS now provides unprecedented metallicities, of much higher quality than virtually all previous estimates,
on a homogeneous scale, and with a relatively large number of high probability members, allowing a substantial improvement in our knowledge of these fascinating, but until now very poorly
studied objects, and of the B/D GC systems in general. A recent complementary step in the same direction was undertaken by Geisler et al. (2023 - hereafter G23) who obtained metallicities and velocities from low resolution Ca triplet (CaT) VLT spectra for 12 B/D GCs, including six of the CAPOS sample, and we will compare these two studies in detail, in both mean metallicity and RV.

We address each of our clusters in turn. First, we note that
none of them show strong evidence for internal metallicity variations.
The sample size in some clusters is relatively small,
from 2-9  members with $SNR>70$, limiting the statistics, 
but four clusters have 10-13 stars each and two have a very large number of high quality members.
In fact, these two clusters, NGC 6273 and 6656, have the two largest standard deviations in metallicity. But this could be explained by either of two possibilities: one is that these are among the most metal-poor clusters in our sample. Indeed, of the four clusters with $[Fe/H]<-1.45$, the mean standard deviation of the metallicity determination is 0.093, twice as large as the mean error for the more metal-rich GCs (0.047). ASPCAP is known to have a more difficult time deriving accurate metallicities for such metal-poor giants, given the weaker lines (Nidever et al. 2020). A second possibility 
is that these two GCs may actually be among the select few that do indeed possess intrinsic metallicity spreads. The history of NGC 6656 is particularly rich and colorful in this regard, with a longstanding debate in the literature about this possibility (Norris and Freeman 1983, Mucciarelli et al. 2015).
Our exquisite APOGEE database is the best available to definitively decide on this issue, but details will be left to a future paper. Suffice it to say here that our spread of 0.10 dex is very similar to the scatter of 0.092 found by Meszaros et al. (2020) in their BACCHUS analysis of 20 APOGEE (non-CAPOS) stars in NGC 6656, in which they do not find any significant intrinsic metallicity variation. Similarly, NGC 6273 might be the remnant nuclear star cluster of a nucleated dwarf galaxy a la  $\omega$ Centauri (Pfeffer et al. 2021) and thus may also possess a small intrinsic metallicity spread (Johnson et al. 2015, 2017).  But such spreads in  the other GCs are essentially ruled out, certainly in the clusters with more than 5 stars, which is most of our sample.

The historic bible of Galactic GC properties, including metallicity,
is H10, so we begin with his values and references and also include more
recent studies. For the H10 error, we give the value from 
Carretta et al. (2009) if his value is based on their compilation.
For the clusters included in G21, we summarize the most relevant studies. The recent Ca triplet investigation by G23 also presents detailed metallicity determinations for many of our sample, and again here we only present the most relevant studies. We will also compare with the S24 value from the APOGEE VAC catalog. We also compare to the results for those clusters that have already been the subject of individual investigations of the CAPOS data via BACCHUS. Finally, we cite any high-resolution studies.

{\bf NGC 6273:}
Our metallicity value is -1.75$\pm0.11$ (the standard deviation is quoted here in all cases for our mean), from 62 members with SNR$>$70, our second-largest sample, including 23 1P stars, and is one of the most metal-poor in CAPOS.  The H10 [Fe/H] value for this cluster is -1.74$\pm 0.07$, based on a variety of non-high resolution spectroscopic methods of reasonable weight,  as assessed by him, placed on the Caretta et al. (2009) metallicity scale, with a reasonably low E(B-V)=0.38 for a BGC. 
This cluster was not included in G21 or G23. 
S24 find -1.71, a slightly larger value than ours, as expected from their not accounting for the spuriously high metallicity of 2P stars. 
High resolution spectra apart from APOGEE have been obtained by Johnson et al. (2015, 2017), and Bailin (2019) gives their mean value as -1.612$\pm 0.022$, but with a large intrinsic spread of 0.16 dex. All values for the mean metallicity are in very good agreement. 

{\bf NGC 6293:}
Our metallicity value is -2.12$\pm0.08$, from 13 members with SNR$>$70, including 5 1P stars. This is the most metal-poor CAPOS cluster. The H10 [Fe/H] value for this cluster is -1.99$\pm 0.14$, based on a variety of non-high resolution spectroscopic methods of high weight placed on the Caretta et al. (2009) metallicity scale, with a reasonably low E(B-V)=0.36. This cluster was not included in G21 or G23. 
S24 find -2.09, a slightly larger metallicity than ours, as expected. High resolution spectra apart from APOGEE have not been published to our knowledge.
All values are in  good agreement. 

{\bf NGC 6304:}
Our metallicity value is -0.49$\pm0.06$, from 12 members with SNR$>$70, including 4 1P stars. This is the most metal-rich CAPOS cluster. The H10 [Fe/H] value for this cluster is -0.45$\pm 0.07$, based on a variety of non-high resolution spectroscopic methods of reasonable weight placed on the Caretta et al. (2009) metallicity scale, with an  intermediate E(B-V)=0.54. This cluster was not included in G21 or G23. 
S24 find -0.48, very slightly larger than ours, as expected from the negligible metallicity effect on 2P stars for this very metal-rich GC. 
A BACCHUS analysis of our data yields -0.45 $\pm $0.08 (Montecinos et al. in prep.).
High resolution spectra apart from APOGEE have not been published to our knowledge.
All values are in excellent agreement. 

{\bf NGC 6316:}
Our metallicity value is -0.83$\pm0.05$, from 6 members with SNR$>$70, including 3 1P stars, making this an intermediate metallicity cluster. The H10 [Fe/H] value for this cluster is -0.45$\pm 0.14$, based on a variety of non-high resolution spectroscopic methods of high weight placed on the Caretta et al. (2009) metallicity scale, with an  intermediate E(B-V)=0.54. This cluster was not included in G21 or G23. 
S24 find -0.77, slightly larger than our value, as expected. A BACCHUS analysis of our data yields $-0.87\pm 0.08$     (Frelijj et al. 2025). 
High resolution spectra apart from APOGEE have not been published to our knowledge.
Our value is 0.38 dex lower than the H10 value, pointing to the difficulty of obtaining accurate metallicities for such GCs. 

{\bf Terzan 2:}
We find a mean metallicity value of -0.88$\pm0.02$, from only 3 members with SNR$>$70  (ASPCAP does not provide abundances for 1 star), including 2 1P stars. This is another intermediate metallicity cluster. The H10 [Fe/H] value for this cluster is -0.69, based on low resolution near-IR spectra. Terzan 2 is very heavily reddened, with E(B-V)=1.87. G21 give [Fe/H]=-0.85$\pm0.04$, while G23 find -0.54$\pm0.10$ from CaT spectra of 8 members, similar to the value of -0.42$\pm0.18$ derived by Vasquez et al. (2018) from the same technique.
S24 find -0.86, slightly larger than our value, as expected. 
 Our CAPOS BACCHUS study yields -0.84 $\pm$ 0.04 (Uribe et al. 2025), which includes the 4th member for which ASPCAP did not provide abundances.
High resolution spectra apart from APOGEE have not been published to our knowledge.
There is a range of almost 0.5 dex in these values, with ours being the lowest, again pointing to the difficulty of obtaining accurate metallicities for such highly obscured GCs and such small samples. 

{\bf Terzan 4:}
Our metallicity value is -1.41$\pm0.04$, from only 3 members with SNR$>$70, including 2 1P stars. This is a metal-poor cluster. The H10 [Fe/H] value for this cluster is -1.41, based on low resolution near-IR spectra. Terzan 4 is very heavily reddened, with E(B-V)=2.0. G21 give [Fe/H]=-1.40$\pm0.05$. G23 did not observe this cluster.
S24 find -1.38, slightly larger than our value, as expected. Our CAPOS BACCHUS study yields -1.42 $\pm$ 0.08 (Sepulveda-Lopez et al. in prep.)
Origlia \& Rich (2004) derive -1.60 based on
high-resolution near-IR spectra of four stars using the Keck
NIRSPEC instrument.
Our mean metallicity is in perfect agreement with H10's,
undoubtedly a case of serendipity given its highly obscured nature and our small sample size. It is also in excellent accord with G21 but almost 0.2 dex higher than the Origlia \& Rich value.

{\bf HP1:}
We derive a mean metallicity value of -1.23$\pm0.07$, from 10 members with SNR$>$70, including 2 1P stars. This is another metal-poor cluster. The H10 [Fe/H] value for this cluster is -1.00, based on low resolution near-IR spectra as well as high-resolution
spectra of two stars (Barbuy et al. 2006). HP1 is highly reddened, with E(B-V)=1.12. G21 find [Fe/H]=-1.20$\pm0.10$, while G23 did not observe this cluster.
S24 find -1.21, slightly larger than our value, as expected. A BACCHUS analysis of our data yields $-1.15\pm 0.08$     (Henao et al. 2025).
VLT UVES studies have been carried out by Barbuy et al. (2006, 2016). In the latter paper, they combine
their sample for a total of 8 stars, deriving a mean metallicity of -1.06$\pm 0.15$.
All values are in good agreement, although our value is the lowest. 

{\bf FSR 1758:}
Our metallicity value is -1.48$\pm0.08$, from 9 members with SNR$>$70, including 2 1P stars. This also is a metal-poor cluster. 
H10 does not include this object in his catalog as it was undiscovered at the time. It has a fairly high E(B-V)=0.9. 
This cluster was not included in G21 or G23. Romero-Colmenares et al. (2021) analyzed the CAPOS spectra but using the BACCHUS package, deriving a mean metallicity of  -1.36 $\pm $0.08.
S24 find -1.42, a slightly larger value than ours, as expected. 
High resolution spectra apart from APOGEE have been obtained by Villanova et al. (2019),
yielding a mean metallicity of -1.58$\pm 0.03$, in good agreement with our value. 

{\bf NGC 6380:}
We find a mean [Fe/H] = -0.90$\pm0.02$, from 4 members with SNR$>$70, including 3 1P stars. This is an intermediate metallicity cluster. The H10 [Fe/H] value for this cluster is -0.75$\pm 0.09$, based on a variety of non-high resolution spectroscopic methods of low weight placed on the Caretta et al. (2009) metallicity scale, with a high E(B-V)=1.17. This cluster was not included in G21 or G23. 
Fernandez-Trincado et al. (2021a) analyzed only 2 of the CAPOS stars plus a number of additional members observed by another APOGEE program and used the BACCHUS package, deriving a mean metallicity of  -0.80 $\pm$ 0.04, with the discrepancy presumably due in part to the different samples.
S24 find -0.78, which is significantly higher than our value. As discussed above, NGC 3680 is a clear example of how S24 included a large number of field stars, which fail one or several of our membership criteria. Figure \ref{N6380Schiavon} shows that many of the field stars they included are more metal-rich than the true cluster members, including one star at solar metallicity, 0.8 dex higher than our most metal-rich member, leading to the substantial metallicity discrepancy. To our knowledge, 
high resolution spectra apart from APOGEE have not been published.
Our value is is reasonable agreement with H10,
but in mediocre agreement with S24, which points to the critical importance of avoiding interloping field stars that can adversely affect cluster means. 

{\bf Ton 2:}
We obtain a mean metallicity of -0.73$\pm0.03$ from 6 members with SNR$>$70, including 3 1P stars. This is another intermediate metallicity cluster. The H10 [Fe/H] value for this cluster is -0.70, based on a variety of non-high resolution spectroscopic methods of very low weight placed on the Caretta et al. (2009) metallicity scale, with a high E(B-V)=1.24. This cluster was not included in G21, while G23 derived -0.57$\pm0.13$ from 16 CaT members. Vasquez et al. (2018) found a value of -0.26$\pm0.15$ from the same technique.
Fernandez-Trincado et al. (2022) analyzed all of the CAPOS spectra but using the BACCHUS package, and derived a mean metallicity of -0.70$\pm0.05$.
S24 find -0.74, slightly lower than our value because they include a number of members observed by APOGEE but not by CAPOS, in addition to also including one very metal-poor field star as well as neglecting the correction for 2P stars. As far as we know,
high resolution spectra apart from APOGEE have not been published heretofore.
Our metallicity is is excellent agreement with that of H10, in reasonable accord with the G23 CaT value, but in very poor agreement with the Vasquez CaT value.

{\bf Terzan 9:}
Our metallicity is -1.42$\pm0.04$, from 9 members with SNR$>$70, including 4 1P stars, making this a low metallicity cluster. The H10 [Fe/H] value for this cluster is -1.05, but note that Carretta et al. (2009) give -2.07$\pm0.09$. Terzan 9 is very heavily reddened, with E(B-V)=1.76. G21 give [Fe/H]=-1.40$\pm0.07$, while G23 find -1.15$\pm0.12$ from CaT spectra of 14 members, similar to the value of -1.08$\pm0.14$ derived by Vasquez et al. (2018) from the same technique. As expected,
S24's value of -1.36 is somewhat larger than our value. No other
high resolution spectra apart from APOGEE exist.
There is a range of over 1 dex in these values, with ours being near the middle, and the 2 CaT values substantially higher.

{\bf Djorg 2:} 
Our metallicity value is -1.14$\pm0.04$, from 6 members with SNR$>$70, including 3 1P stars. This is an intermediate metallicity cluster. The H10 [Fe/H] value for this cluster is -0.65, with this value being based on measurements of very low weight. Djorg 2 has a fairly large  reddening, with E(B-V)=0.94. G21 give [Fe/H]=-1.07$\pm0.09$, 
using an additional member that we do not include, while G23 find -0.67$\pm0.10$ from CaT spectra of only 2 members, 
substantially higher than the value of -0.97$\pm0.13$ derived by Vasquez et al. (2018) from the same technique. Once again, as expected, the 
S24 value of -1.07 is somewhat larger than our value.  Our CAPOS BACCHUS analysis finds -1.04 $\pm$ 0.06 (Pino-Zuniga et al. 2025).
Kunder \& Butler (2020) use our CAPOS sample and ASPCAP to derive -1.05$\pm0.08$, again ignoring any effect on 2P stars.
High resolution spectra apart from APOGEE have not been published to our knowledge.
Our value is in poor agreement with those of H10 and G23, but in reasonable accord with that of Vasquez et al. (2018).

{\bf NGC 6540:} 
Our derived metallicity is -1.09$\pm0.06$, from 4 members with SNR$>$70, including only 1 1P star. This is an intermediate metallicity cluster. The H10 [Fe/H] value for this cluster is -1.35, with low weight based on high resolution but low SNR optical spectra. NGC 6540 has an intermediate reddening, with E(B-V)=0.66. G21 give [Fe/H]=-1.06$\pm0.06$,  while G23 find -1.04$\pm0.14$ from CaT spectra of 5 members.
S24 find -1.02, somewhat larger than our value, as expected. 
High resolution spectra apart from APOGEE have not been published to our knowledge.
Our value is in poor agreement with H10 but very good accord with G23.

{\bf NGC 6558:}
Our metallicity for this cluster is -1.15$\pm0.03$, from 4 members with SNR$>$70, including only 1 1P star. This is an intermediate metallicity cluster. The H10 [Fe/H] value is -1.32$\pm 0.14$, based on a variety of methods of high weight placed on the Caretta et al. (2009) metallicity scale, including high resolution optical spectra for 5 stars from Barbuy et al. (2007), who find -0.97$\pm0.15$. NGC 6558 has a relatively low E(B-V)=0.44. This cluster was not included in G21 or G23. 
S24 find -0.99, a significantly larger value than ours, with a difference even larger than expected from their not accounting
for the spuriously high metallicity of 2P stars. The additional discrepancy comes mostly from their inclusion of 1 PM member which however is 0.5 dex more metal-rich than our most metal-rich member.
 The discrepancy appears to be due to the very low SNR (10) of this star, yielding a poor ASPCAP metallicity.
Gonzalez-Diaz et al. (2023) use our CAPOS sample and BACCHUS to derive -1.15$\pm0.08$.
Barbuy et al.'s  (2018) assessment of -1.17$\pm0.10$ from high-resolution spectra of 4 RGB stars is almost identical to ours while H10 is significantly lower.

{\bf NGC 6569:}
For this cluster we find a mean [Fe/H] = -1.03$\pm0.05$, from 7 members with SNR$>$70, including 4 1P stars, so it is another intermediate metallicity cluster.  The H10 [Fe/H] value is -0.76$\pm 0.14$, based on a variety of non-high resolution spectroscopic methods of high weight placed on the Caretta et al. (2009) metallicity scale, with an intermediate E(B-V)=0.53. 
This cluster was not included in G21 or G23. 
S24 find -0.92, a significantly larger value than our's, with a difference even larger than expected from their not accounting for the spuriously high metallicity of 2P stars. The additional discrepancy comes from their inclusion of 2 PM non-members that are also as much as 1.2 dex more metal-rich than our most metal-rich member, once again demonstrating the danger of not eliminating field contaminants.
A BACCHUS analysis of our data yields $-0.91\pm 0.06$    (Barrera et al. 2025).
High resolution spectra apart from APOGEE have been obtained by Johnson et al. (2018), while Bailin (2019) gives the mean value as -0.867$\pm 0.014$.  Agreement with H10 is mediocre
but reasonable with Johnson et al. (2018). 

{\bf NGC 6642:}
Our metallicity value is -1.11$\pm0.04$, from 10 members with SNR$>$70, including 6 1P stars. This is yet another intermediate metallicity cluster. The H10 [Fe/H] value is -1.26$\pm0.14$, with very high weight based on a variety of methods including  low resolution optical spectra. NGC 6642 has  relatively low reddening, with E(B-V)=0.40. G21 give [Fe/H]=-1.11$\pm0.04$, using a much smaller APOGEE sample than now available to us,  while G23 find -1.11$\pm0.24$ from CaT spectra of 19 members.
S24 find -1.09, slightly higher than our value, as expected. 
High resolution spectra apart from APOGEE have not been published to our knowledge.
Our value is in good agreement with H10 and perfect accord with G23.

{\bf NGC 6656:}
This is a very nearby and well-studied cluster, and contains by far our largest sample of CAPOS members. APOGEE also obtained a large number of additional stars for a different program, which we do not include here.
Our metallicity value is -1.75$\pm0.10$, from 125 members with SNR$>$70, including 58 1P stars. This is a very metal-poor cluster.  The H10 [Fe/H] value is -1.70$\pm 0.08$, based on a variety of methods including multiple high resolution optical spectroscopic studies of extremely high weight placed on the Carretta et al. (2009) metallicity scale, with a reasonably low E(B-V)=0.34. 
This cluster was not included in G21 or G23. 
S24 find -1.70,  slightly larger than our value, 
utilizing a large number of the additional APOGEE stars not included in CAPOS. 

A sample of these additional APOGEE stars were analyzed by Meszaros et al. (2020), finding -1.524$\pm0.112$
from 20 stars with SNR$>$70. 
Correcting for internal error in the abundance determination, their overall estimate of the intrinsic metallicity error is 0.092. 
They did not confirm a metallicity variation in NGC 6656, which is controversial. 
Other recent high resolution spectra apart from APOGEE have been obtained by Marino et al. (2011) and Mucciarelli et al. (2015), and Bailin (2019) gives their mean value as -1.803$\pm 0.015$, again with no strong evidence for an intrinsic metallicity dispersion.  Although our error is relatively high, as discussed above this is expected for such a metal-poor cluster, and we reinforce the likelihood that NGC 6656 does not possess a large intrinsic metallicity spread, if any. 
Also note that we have not corrected our estimate of the intrinsic metallicity error for the internal error in abundance determination. Munoz et al. (2021) also concluded that NGC 6656 likely did not have a strong, if any, intrinsic metallicity spread.

Agreement of our mean metallicity with that from H10 is very good, presumably because his value involves many high resolution studies, as is agreement with recent non-APOGEE high resolution studies. The  reason for the  offset with Meszaros et al. is uncertain.

{\bf NGC 6717:} 
Our metallicity value is -1.17$\pm0.05$, from only 2 members with SNR$>$70, both 1P stars. This is yet another intermediate  metallicity cluster. The H10 [Fe/H] value for this cluster is -1.26$\pm 0.07$, based on a variety of non-high resolution spectroscopic methods of medium weight placed on the Caretta et al. (2009) metallicity scale, with a reasonably low E(B-V)=0.22. This cluster was not included in G21 or G23. 
S24 find -1.12, a slightly larger value than ours.  However, the cause is not due to 2P stars as there are none observed in this cluster, but instead due to an outlier beyond the tidal radius in  our analysis which was included by S24 which has a higher metallicity than our members. High resolution spectra apart from APOGEE have not been published to our knowledge.
Our value is in good agreement with H10. 

In summary, our mean metallicities are generally in good to very good agreement with previous high quality spectroscopic studies but these are limited due to the high extinction that most of our sample suffers.  The metallicity correction we make for 2P stars can make a substantial difference in the final mean metallicity derived from previous ASPCAP studies like S24.
Agreement with previous lower quality spectroscopic or photometric determinations, particularly those based on optical data, is more 
variable, as expected given the less precise data. We emphasize that the near-IR, high resolution and high SNR observations, relatively large sample sizes and excellent membership probabilities, as well as homogeneous abundance derivations make our CAPOS results unparalleled.

 Nevertheless, clearly any high resolution abundance analysis depends on many factors such as the atmospheric parameters used. There are now BACCHUS analyses for 11 CAPOS GCs where we can compare results. Note that the samples are generally the same but occasionally slightly different, as noted above, and that the BACCHUS analyses use atmospheric parameters derived from photometry, unlike the spectroscopic ones in ASPCAP. For this sample, we find that the ASPCAP mean [Fe/H] is 0.05 $\pm $ 0.06 dex lower than the BACCHUS mean. We consider this difference to be within the respective errors.

\begin{table*}
\captionsetup{justification=centering, singlelinecheck=false}
\centering
\caption{Mean cluster metallicity, [$\alpha$/Fe] and radial velocity for members.}
\label{meanvalues}
\begin{threeparttable}
\begin{tabular}{l l c r c}
\hline\hline
\multicolumn{1}{c}{Cluster ID} &
\multicolumn{1}{c}{[Fe/H] (dex)} &
\multicolumn{1}{c}{[$\alpha$/Fe] (dex)} &
\multicolumn{1}{c}{$V_{r}$ (km s$^{-1}$)} &
\multicolumn{1}{c}{N$_{1\mathrm{P}}$} \\
\hline
NGC 6273 &  $-$1.75$\pm${ 0.11} & 0.22$\pm$0.07 & 144.7$\pm$7.2 &   23 \\
 NGC 6293 &  $-$2.12$\pm${ 0.08} & 0.29$\pm$0.08 & $-$142.9$\pm$5.6 &   5 \\
  NGC 6304 &  $-$0.49$\pm${ 0.06} & 0.18$\pm$0.04 & $-$109.6$\pm$5.2 &    4 \\
  NGC 6316 &  $-$0.83$\pm${ 0.05} & 0.23$\pm$0.04 & 100.7$\pm$3.6 &   3 \\
  Terzan 2 &  $-$0.88$\pm${ 0.02} & 0.25$\pm$0.02 & 134.1$\pm$1.1 &   2 \\
  Terzan 4 & $-$1.41$\pm${ 0.04} & 0.22$\pm$0.04 & $-$47.7$\pm$3.5 &  2 \\
  HP 1 &  $-$1.23$\pm${ 0.07} & 0.23$\pm$0.04 & 40.1$\pm$4.2 &   2 \\
  FSR 1758 &  $-$1.48$\pm${ 0.08} & 0.25$\pm$0.05 & 224.6$\pm$2.9 &   2 \\
  NGC 6380 &  $-$0.90$\pm${ 0.02} & 0.20$\pm$0.09 & 1.7$\pm$5.5 &    3 \\
  Ton 2 &  $-$0.73$\pm${ 0.03} & 0.26$\pm$0.05 & $-$179.0$\pm$3.7 &    3 \\
  Terzan 9 & $-$1.42$\pm${ 0.04} & 0.24$\pm$0.04 & 70.4$\pm$5.3 &   4 \\
  Djorg 2 &  $-$1.14$\pm${ 0.04} & 0.24$\pm$0.03 & $-$151.9$\pm$1.2 &  3 \\
  NGC 6540 & $-$1.09$\pm${ 0.06} & 0.23$\pm$0.05 & $-$14.3$\pm$1.1  &  1 \\
  NGC 6558 &  $-$1.15$\pm${ 0.03} & 0.22$\pm$0.03 & $-$192.6$\pm$1.2 &    1 \\
  NGC 6569 &  $-$1.04$\pm${ 0.05} & 0.29$\pm$0.08 & $-$49.7$\pm$3.5 &   5 \\
  NGC 6642 &  $-$1.11$\pm${ 0.04} & 0.31$\pm$0.05 & $-$55.5$\pm$2.4 &   6 \\
  NGC 6656 &  $-$1.75$\pm${ 0.10} & 0.29$\pm$0.06 & $-$147.4$\pm$6.2 &    59 \\
  NGC 6717 &  $-$1.17$\pm${ 0.05} & 0.25$\pm$0.01 & 28.3$\pm$1.4 &   2 \\

\hline
\end{tabular}
\begin{tablenotes}{\small
\item  $[\ensuremath{\alpha}/Fe]$ is the mean [Si/Fe] abundance including 1P and 2P stars.}
\end{tablenotes}
\end{threeparttable}
\end{table*}

\subsection{The $\alpha$-
elements}
The $\alpha$-elements play a crucial role in divulging the chemical evolution
of a system, in particular the past rate of star formation, as
well as information on the IMF. In concert with the metallicity,
they reveal the onset of the dominance of SNe Ia over SNe II. In
addition, a detailed knowledge of the [$\alpha$/Fe] abundance ratio is critical to derive an accurate age estimate from a GC deep CMD.

As noted in G21, there is some freedom as to which particular element or combination thereof to use to best represent the $\alpha$ abundance from APOGEE data. All four different
possibilities investigated here – global $\alpha$, Mg, Si,
and Ca (Table 5) - are all very well-determined in all of our sample,
with typical errors of the mean $\sim$0.05 dex, with
the exception of global $\alpha$ and especially Ca in the most metal-poor clusters, which possess large spreads.
All four mean values
for a given GC are also in  good accord, with the different
means falling within about 0.1 dex.
There are some small systematic offsets – on average for our
sample, the highest mean cluster abundance is that of [Mg/Fe],
followed by [$\alpha$/Fe] (0.03 dex lower), [Si/Fe] (0.06 dex lower), 
and finally [Ca/Fe] (0.11 dex lower). These offsets are quite similar to what we found in G21 for a subset of our clusters, as expected. It is likely that the differences are real and that small real mean abundance differences exist among these elements, thus exacerbating  selecting only one of them to represent all of the  $\alpha$ elements. However, we have opted for Si as the best representative since it is virtually unaffected by 2P atmospheric parameter issues  or abundance variations, as opposed to our selection of global $\alpha$ in G21. We therefore show the mean [Si/Fe] abundance in Table 4 as our best representative of [$\alpha$/Fe]. 

We first compare our [Si/Fe] abundances with previous derivations of this ratio using CAPOS data (in some cases supplemented with a small number of additional stars from the main APOGEE survey) and the BACCHUS code, using photometric atmospheric parameters derived in a homogeneous manner.  Montecinos et al. (in prep.) found a mean [Si/Fe] = 0.31$\pm$0.10 in NGC 6304, substantially higher than our ASPCAP value of 0.18$\pm $0.04. Frelijj et al. (2025) used our CAPOS sample of NGC 6316, supplemented by 2 additional stars, and their BACCHUS analysis yielded 
[Si/Fe] = 0.36$\pm$0.05, substantially higher than our ASPCAP value of 0.23$\pm$0.04.  The analysis of Terzan 2 by Uribe et al. (2025) found a mean [Si/Fe] 0.08 dex higher than our value., while that of Terzan 4 by Sepulveda-Lopez (in prep.) is 0.19 dex higher.
Henao et al's. (2025) analysis of our CAPOS sample for HP 1 gave 
[Si/Fe] = 0.45$\pm$0.07, substantially higher than our value of 0.23$\pm$0.04.
Romero-Colmenares et al. (2021) used basically our CAPOS sample 
(including 2 extra stars) for FSR 1758 to derive [Si/Fe] =  0.33$\pm$0.05, somewhat higher than our value of 0.25$\pm$0.05.
Fernandez-Trincado et al. (2021a) analyzed the CAPOS spectra for NGC 6380 plus a number of additional members, deriving [Si/Fe] = 0.37$\pm$0.04, substantially higher than  our value of 0.20$\pm$0.09. 
Fernandez-Trincado et al. (2022) analyzed the CAPOS spectra for Ton 2 plus one additional CAPOS member with SNR=63, slightly below our limit of 70, deriving [Si/Fe] = 0.33$\pm$0.06, somewhat higher than  our value of 0.26$\pm$0.05.  The recent study of Djorg 2 by Pino-Zuniga et al. (2025) yielded a BACCHUS value of 0.38$\pm$0.05, compared to the ASPCAP mean of 0.24$\pm$0.03. Gonzalez-Diaz et al. (2023) derived a mean value only 0.02 dex higher than our value of 0.22$\pm$0.03 for NGC 6558. Finally,
Barrera et al. (2025) basically used the CAPOS sample for NGC 6569
and found [Si/Fe] =  0.35$\pm$0.07, slightly higher than  our value of 0.29$\pm$0.08.

Clearly, BACCHUS mean values are  generally significantly higher than the corresponding ASPCAP means, indicating likely systematic effects between the two techniques. There are small differences in the samples for some of the clusters but the standard deviations in all clusters are quite small compared to the differences so a few additional stars should not effect the means very much. For these  11 GCs, we find a mean difference of [Si/Fe] (BACCHUS - ASPCAP) =   +0.12$\pm 0.06$, which we consider larger than the respective errors.
We are unsure of the reason for this discrepancy. They do use distinct atmospheric parameters but one needs to explore if there are systematic differences in these.  We have begun to investigate this behavior. More details will be given in Montecinos et al. (in prep.), but a preliminary analysis indicates that, for good Si lines, BACCHUS generally gives a slightly better fit to the observed spectrum than ASPCAP, using the same atmospheric parameters, suggesting that the BACCHUS Si abundances are slightly preferable.

The mean [Si/Fe] for all CAPOS GCs is 0.24 ($\sigma $ = 0.03), while for S24 the mean for the same sample is 0.23 ($\sigma $ = 0.05).
We note that Kunder \& Butler (2020) derive [Si/Fe] = 0.25$\pm$0.08
for Djorg 2, 0.01 dex higher than our value, from our CAPOS data but also including a star that we discarded as a non-member.

We next compare our [Si/Fe] abundances with previous measurements based on other high resolution spectra.
Johnson et al. (2015, 2017) investigated a large number of stars in NGC 6273 and found evidence for three populations: metal-poor with mean [Fe/H] = -1.77$\pm$0.08 and [Si/Fe] = 0.35$\pm$0.10, metal-intermediate with 
[Fe/H] = -1.51$\pm$0.07 and [Si/Fe] = 0.23$\pm$0.15 and metal-rich with [Fe/H] = -1.22$\pm$0.09 and [Si/Fe] = 0.11$\pm$0.21. We find no strong evidence for a spread in either Fe or Si, with a mean [Si/Fe] = 0.22$\pm$0.07, in the middle of the range found by Johnson et al. (2017). 
Origlia \& Rich (2004) find [Si/Fe] = 0.55 for Terzan 4, much higher than our value of 0.22$\pm$0.04 and indeed exceeding essentially all high resolution estimates of GC Si abundances.
Barbuy et al (2016) determine [Si/Fe] = 0.27 for HP1, within the errors of our value of 0.23$\pm$0.04.
Barbuy et al. (2007) derived 0.23$\pm$0.03 for NGC 6558, virtually identical to our value of 0.22$\pm$0.03.
Johnson et al. (2018) derived 0.34$\pm$0.09  for NGC 6569, in good agreement with our mean of 0.29$\pm$0.08.
Marino et al. (2011) derived 0.44$\pm$0.06  for NGC 6656, in poor agreement with our mean of 0.29$\pm$0.06.

In conclusion, our values are in general good agreement with the limited literature available, although the systematic offset with CAPOS BACCHUS values is disconcerting.

\subsection{Radial velocity}
Radial velocities are powerful membership criteria, provide
insight into internal cluster dynamics, and are key ingredients
to derive the cluster orbit and thus constrain its origin. Here we compare our mean RVs to those in Baumgardt et al. (2019 - hereafter B19), as the best recent global compilation, replacing H10, as well as to G23 for clusters in common. 

In  general, the agreement with B19 is very good to excellent but there are some exceptions and two very notable ones. The differences are generally within a few km/s, with five clusters exceeding 5km/s, but most problematic are Terzan 9 and NGC 6642, whose differences are 40 and 22 km/s, respectively. If we consider all clusters (excluding FSR 1758, which is not included in B19), the mean difference (in the sense us - B19) is 1.5$\pm$12.1 km/s, while excluding Terzan 9 and NGC 6642 yields values of only 0.5$\pm$4.3 km/s, which are well within the errors. The same values with respect to G23 are 2.9$\pm$6.2 km/s, also within the respective errors (the G23 CaT RV errors are larger). Moreover, the differences with G23 for Terzan 9 and NGC 6642 are only 0.3 and 7.2 km/s. We agree with G23 that the
B19 RVs for these two clusters are highly suspect. The RV differences for these two clusters are significant and important, and for example would affect the orbital determination as well as the classification to a specific Galactic component. A check on these two clusters with the new RVs from Gaia DR3 gives mean values of 66.7$\pm 9.6$ km/s for Terzan 9 and 
-53.7$\pm 5.0$ km/s for NGC 6642, in excellent agreement with our findings.
We note that of the seven clusters with differences with B19 $>$5 km/s, five of them are the most reddened in our sample, including Terzan 9 (but not NGC 6642). Many of the measurements in the B19 compilation come from optical observations, so the large extinction and thus limited SNR in the optical for such clusters may be the cause of many if not all the discrepancies.
Finally, our mean RV is within 2 km/s of that of Villanova et al (2019) for FSR 1758.

\section{Other elements}

\subsection{Light elements - MP}
Of the traditional light elements that APOGEE measures well, all of them are also subject to intrinsic cluster variations due to MP. All of our GCs  show MP except NGC 6717, as seen in Figure \ref{FevsN}
 (where we have implicitly assumed that at least N abundances are indeed well measured by ASPCAP). We have shown that ASPCAP has issues with the atmospheric parameters and derived abundances for 2P stars for at least some other elements, although we have also shown that Si is virtually unaffected by this. Additionally, Si is also the light element least affected by MP. Therefore we only 
present mean abundances for all of the other light elements here, in which we use only 1P stars to derive these mean values (Table 5), with the exception of Si, but we defer further  discussion of Si until the next section,
and postpone analysis of the other light elements and MP in general to future papers in which we derive abundances from BACCHUS, as done in CAPOS papers II - VII (Romero-Colmenares et al. 2021, Fernandez-Trincado et al. 2022, Gonzalez-Diaz et al. 2023, Henao et al. 2025, Barrera et al. 2025, Frelijj et al. 2025, respectively)
for several clusters already.
We will also not discuss here the K or Ca results (also given in Table 5) because of their relatively large errors, especially for metal-poor clusters.

\subsection{Fe-peak elements}
ASPCAP produces
abundances of the Fe-peak elements V, Cr, Mn, Co, Ni, and Cu.
However, the reliability of these abundances can vary with the
number of lines, their strengths, limitations of the pipeline, etc.
As discussed in G21, Jonsson et al. (2020) have undertaken a painstaking assessment of the reliability of ASPCAP abundances. They rate Mn and Ni as the best Fe-peak species in both accuracy and precision, and find problems with
V, Cr, Co, and Cu. Therefore, we abide by their assessment and again
only investigate these two elements here (Table 5).

\begin{figure*}
\centering
\includegraphics[width=14cm]{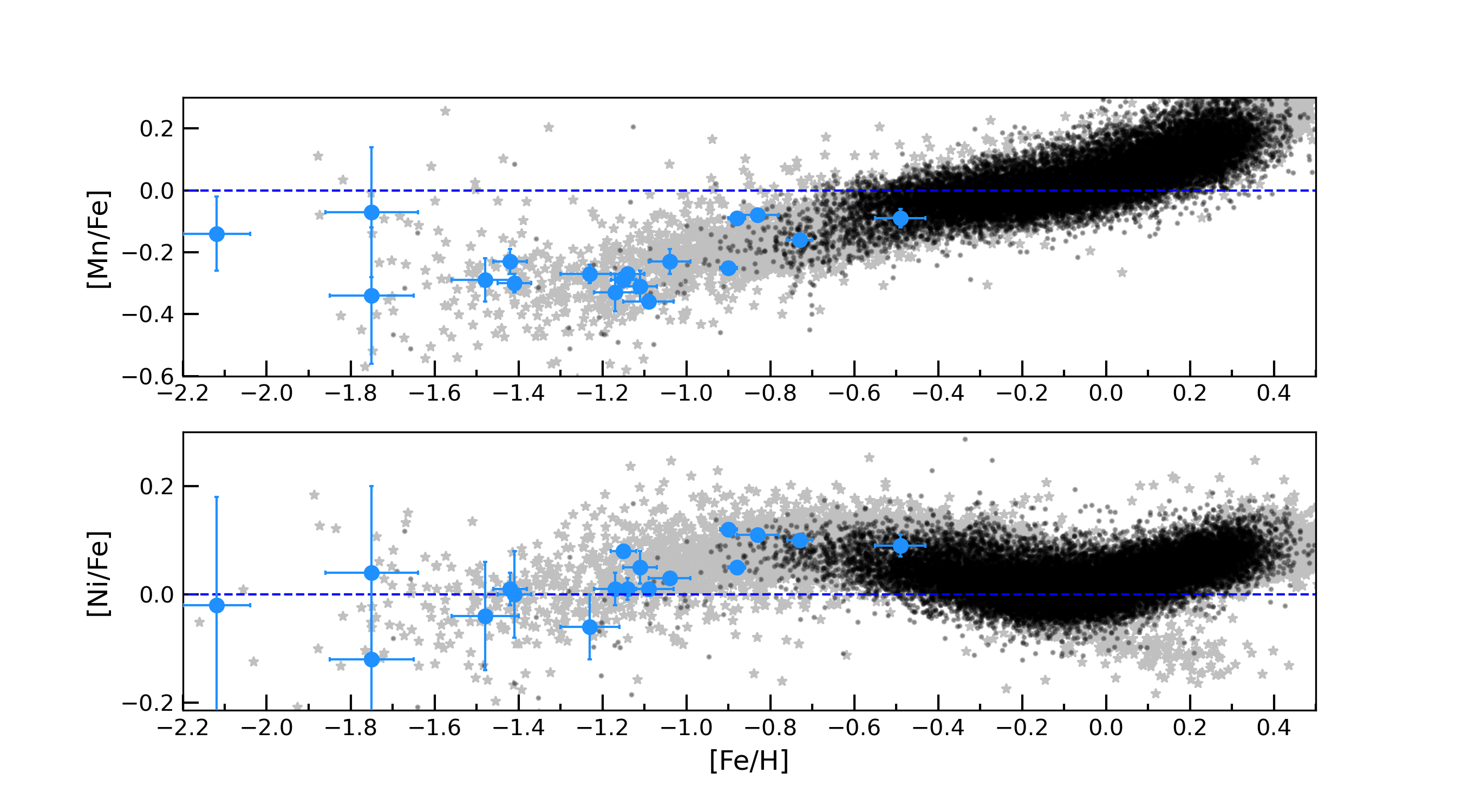}
      \caption{Mean abundance ratio of the Fe-peak elements Mn (top) and Ni (bottom) for each of our clusters (filled blue circles  with error bars), compared with APOGEE bulge stars (asterisks - Rojas-Arriagada et al. 2020) and the general trend of disk stars  (circles -  Hayden  et al. 2015). Our BGCs generally follow the bulge field-star trend but  extend to  lower metallicity.}
    \label{Mn_Ni}
\end{figure*}

The behavior of the mean [Mn/Fe] and [Ni/Fe] versus [Fe/H]
in our clusters is shown in Fig. \ref{Mn_Ni}, and compared with the trends
of APOGEE (ASPCAP) bulge field stars (Rojas-Arriagada et al.
2020), as well as disk field stars ( Hayden et al. 2015, Ernandes et al. 2020). First, notably, our
sample covers a metallicity range so far relatively underexplored
in both species in  both Galactic components,
extending to considerably lower metallicities in general.
Indeed, the
fraction of bulge field stars with metallicities below $\sim -1$ is very
small (Rojas-Arriagada et al. 2020), and although disk stars have
all but disappeared at these low metallicities, these elements are poorly studied. However, our
results lie well within the bulge field star locus at these low
metallicities, so there is good general agreement between our
bulge cluster and field star samples in both Mn and Ni. 

Measurements of Ni abundances independent of APOGEE are sparse for our clusters, while we found no other published Mn abundances. Johnson et al. (2015) derived $<[Ni/Fe]> = -0.02\pm 0.10$ in NGC 6273 from 17 stars, comparable to our value of  $0.04\pm 0.16$, also in good agreement with the mean of $-0.05\pm 0.11$ from Johnson et al. (2017) from an even larger sample. Villanova et al. (2019) derived $<[Ni/Fe]> = -0.09\pm 0.02$ in FSR 1758 from 9 stars, comparable to our value of  $-0.04\pm 0.10$. Johnson et al. (2018) derived $<[Ni/Fe]> = -0.08\pm 0.05$ in NGC 6569 from 19 stars, somewhat lower than  our value of  $0.03\pm 0.01$. 

\begin{table*}
\centering
\begin{threeparttable}[b]
\caption{Mean abundances of First Population stars (Si includes 2P stars as well)}
\label{Element}
\centering
\begin{tabular}{ l@{\hskip 4pt} c@{\hskip 4pt} c@{\hskip 4pt} c@{\hskip 4pt} c@{\hskip 4pt} 
c@{\hskip 4pt} c@{\hskip 4pt} c@{\hskip 4pt} c@{\hskip 4pt} c@{\hskip 4pt} c@{\hskip 4pt} 
c@{\hskip 4pt} c }
\hline 
\hline
Cluster ID & [C/Fe] & [N/Fe] & [O/Fe] & [Na/Fe] & [Mg/Fe] & [Al/Fe]
& [Si/Fe] & [K/Fe] & [Ca/Fe] & [Mn/Fe] & [Ni/Fe] &  $ [\alpha /Fe]$ \\
\hline
NGC 6273   & -0.26      	  &  0.49     	    &  0.23         &    0.09	  	 &  0.21	   &   0.06         &   0.22 & 0.20 & 0.23 & -0.07 & 0.04 &    0.20    \\
          & $\pm$0.29  	  &  $\pm$0.24 	    &  $\pm$0.17     &    $\pm$0.67  	 &  $\pm$0.11	   &   $\pm$0.41     &   $\pm$0.07 & $\pm$0.36 & $\pm$0.27 & $\pm$0.21 & $\pm$0.16 & $\pm$0.10  \\
	  
NGC 6293    & -0.69	  &  0.54           &  0.34          &    0.37       	 &  0.26	   &   -0.04          &   0.29 & -0.07 & -0.07 & -0.14 & -0.02 &  0.26        \\
          & $\pm$0.31  	  &  $\pm$0.13 	    &  $\pm$0.02     &    $\pm$0.17  	 &  $\pm$0.04	   &   $\pm$0.37     &   $\pm$0.08 & $\pm$0.17 & $\pm$0.50 & $\pm$0.12 & $\pm$0.20 &  $\pm 0.08$  \\
	  
NGC 6304    & 0.14       	  &  0.16      	    &  0.31          &    0.13       	 &  0.33       	   &   0.20          &   0.18 & 0.32 & 0.12 & -0.09 & 0.09 &  0.27     \\
          & $\pm$0.03  	  &  $\pm$0.02 	    &  $\pm$0.02     &    $\pm$0.08  	 &  $\pm$0.03	   &   $\pm$0.01     &   $\pm$0.04 & $\pm$0.03 & $\pm$0.01 & $\pm$0.03 & $\pm$0.02 &  $\pm$0.02  \\
	  
NGC 6316   & -0.02	  &  0.33           &  0.25         &   -0.09	  	 &  0.30	   &   0.03          &   0.23 & 0.17 & 0.15 & -0.08 & 0.11 &  0.22     \\ 
          & $\pm$0.04  	  &   $\pm$0.25 	    &  $\pm$0.06     &    $\pm$0.21  	 &  $\pm$0.03	   &   $\pm$0.22     &   $\pm$0.04 & $\pm$0.07 & $\pm$0.07 & - & $\pm$0.00 &  $\pm0.05$  \\
	  
Terzan 2   & -0.05	  &  0.38           &  0.32          &   -0.03	  	 &  0.37       	   &   0.04          &   0.25 & 0.28 & 0.24	& -0.09 & 0.05 &  0.28   \\ 
          & $\pm$0.01  	  &  $\pm$0.24 	    &  $\pm$0.00     &    $\pm$0.04  	 &  $\pm$0.04	   &   $\pm$0.00     &   $\pm$0.02 & $\pm$0.01 & $\pm$0.00 & - & 0.00 &  $\pm0.01$  \\
	  
Terzan 4   & -0.43       	  &  0.42      	    &  0.30          &    0.03       	 &  0.23	   &   -0.11          &   0.22 & 0.23 & 0.19 & -0.30 & 0.00 &  0.26    \\ 
          & $\pm$0.01  	  &  $\pm$0.17 	    &  $\pm$0.01     &    -  	 &  $\pm$0.01	   &   $\pm$0.13     &   $\pm$0.04 & $\pm$0.02 & $\pm$0.02 & $\pm$0.03 & $\pm$0.08 &  $\pm0.01$  \\
	  
HP 1   & -0.31 &  0.29           &  0.30          &    -0.18       	 &  0.27	   &   -0.10          &   0.23 & 0.15 & 0.23  & -0.27 & -0.06 &  0.27    \\
          & $\pm$0.06  	  &  $\pm$0.05 	    &  $\pm$0.00     &    $\pm$0.22  	 &  $\pm$0.03	   &   $\pm$0.04     &   $\pm$0.04 & $\pm$0.14 & $\pm$0.06 & $\pm$0.03 & $\pm$0.06 &  $\pm0.00$\\
	  
FSR 1758    & -0.48       	  &  0.32      	    &  0.42          &    0.29       	 &  0.30       	   &   -0.13          &   0.25 & 0.17 & 0.21  & -0.29 & -0.04 &  0.32     \\
          & $\pm$0.07  	  &  $\pm$0.15 	    &  $\pm$0.01     &    $\pm$0.30  	 &  $\pm$0.06  	   &   $\pm$0.00     &   $\pm$0.05 & $\pm$0.13 & $\pm$0.12 & $\pm$0.07 & $\pm$0.10 &  $\pm0.01$ \\
	  
NGC 6380   & 0.01       	  &  0.41      	    &  0.24          &   -0.06       	 &  0.28       	   &   0.21          &   0.20 & 0.22 & 0.16  & -0.25 & 0.12 &  0.28     \\
          & $\pm$0.06  	  &  $\pm$0.16 	    &  $\pm$0.12     &    $\pm$0.28  	 &  $\pm$0.05  	   &   -     &   $\pm$0.09 & $\pm$0.06 & $\pm$0.08 & - & $\pm$0.01 &  $\pm0.10$  \\
	  
Ton 2   & 0.06      	  &  0.35     	    &  0.29         &    -0.07      	 &  0.32      	   &   0.11         &   0.26 & 0.18 & 0.12  & -0.16 & 0.10 &  0.26    \\
          & $\pm$0.04  	  &  $\pm$0.21 	    &  $\pm$0.06     &    $\pm$0.12  	 &  $\pm$0.05  	   &   $\pm$0.00     &   $\pm$0.05 & $\pm$0.00 & $\pm$0.01 & - & $\pm$0.01 &  $\pm0.05$  \\
	
Terzan 9   & -0.46    	  &  0.38     	    &  0.26         &    -0.15      	 &  0.27      	   &   -0.19         &   0.24 & 0.37 & 0.25   & -0.23 & 0.01 &  0.22   \\
          & $\pm$0.05 	  &  $\pm$0.15 	    &  $\pm$0.02     &    $\pm$0.38  	 &  $\pm$0.05  	   &   $\pm$0.05     &   $\pm$0.04 & $\pm$0.10 & $\pm$0.04 & $\pm$0.04 & $\pm$0.03 &  $\pm0.01$ \\	  

Djorg  2   & -0.22     	  &  0.31     	    &  0.34         &    -0.21      	 &  0.33      	   &   0.02         &   0.24 & 0.22 & 0.20 & -0.27 & 0.01 &  0.30      \\
          & $\pm$0.07  	  &  $\pm$0.03 	    &  $\pm$0.02     &    $\pm$0.08  	 &  $\pm$0.03  	   &   $\pm$0.08     &   $\pm$0.03 & $\pm$0.01 & $\pm$0.06 & $\pm$0.00 & $\pm$0.02 &  $\pm0.02$  \\
          
NGC 6540   & -0.11     	  &  0.13     	    &  0.41         &    0.29      	 &  0.35      	   &   0.04         &   0.23 & 0.28 & 0.26 & -0.36 & 0.01 &  0.32    \\
          & -  	  &  - 	    &  -     &    -  	 &  -  	   &   -     &   $\pm$0.05  & - & - & - & - &  $\pm0.00$\\         

NGC 6558   & -0.27  	  &  0.16     	    &  0.38         &    0.07      	 &  0.31      	   &   -0.01         &   0.22 & 0.34 & 0.26 & -0.29 & 0.08 &  0.28    \\
          & -  	  &  - 	    &  -     &    -  	 &  -  	   &   -     &   $\pm$0.03  & - & - & - & - &  $\pm0.00$\\

NGC 6569   & -0.10     	  &  0.33     	    &  0.30         &    -0.20      	 &  0.31      	   &   0.16         &   0.29 & 0.22 & 0.18 & -0.23 & 0.03 &  0.26      \\
          & $\pm$0.06  	  &  $\pm$0.15 	    &  $\pm$0.11    &    $\pm$0.15 
        &  $\pm$0.05  	   &   $\pm$0.03     &   $\pm$0.08 & $\pm$0.08 & $\pm$0.07 & $\pm$0.04 & $\pm$0.01 &  $\pm0.09$ \\    
          
NGC 6642   & -0.14      	  &  0.25     	    &  0.37         &    0.07      	 &  0.32      	   &   0.15         &   0.31 & 0.22 & 0.22 & -0.31 & 0.05 &  0.31  \\
          & $\pm$0.05  	  &  $\pm$0.06 	    &  $\pm$0.05     &    $\pm$0.34  	 &  $\pm$0.02  	   &   $\pm$0.06     &   $\pm$0.05 & $\pm$0.10 & $\pm$0.03 & $\pm$0.05 & $\pm$0.03 &  $\pm0.02$ \\

NGC 6656   & -0.16      	  &  0.28     	    &  0.34         &    0.23      	 &  0.29      	   &   0.08         &   0.29 & 0.20 & 0.18 & -0.34 & -0.12 &  0.26     \\
          & $\pm$0.34  	  &  $\pm$0.36 	    &  $\pm$0.18     &    $\pm$0.56  	 &  $\pm$0.07  	   &   $\pm$0.38     &   $\pm$0.06 & $\pm$0.37 & $\pm$0.29 & $\pm$0.22 & $\pm$0.17 &  $\pm0.09$ \\

NGC 6717   & -0.27    	  &  0.32     	    &  0.38         &    0.11      	 &  0.32      	   &   -0.13         &   0.25 & 0.24 & 0.19 & -0.33 & 0.01 &  0.28    \\
          & $\pm$0.04  	  &  $\pm$0.04 	    &  $\pm$0.09     &    $\pm$0.21  	 &  $\pm$0.02  	   &   $\pm$0.02     &   $\pm$0.01 & $\pm$0.01 & $\pm$0.02 & $\pm$0.06 & $\pm$0.03 &  $\pm0.01$ \\          

\hline
\end{tabular}
\end{threeparttable}
\end{table*}

\section{Major implications}

\subsection{Nature of our sample: a revised chemo-dynamical classification for Galactic GCs}

With the advent of CAPOS, the number of well-studied GCs towards the GB has increased substantially,
in particular with regards to detailed abundances and velocities. 
A new general study of the BGC  system is warranted, revisiting such major themes as the chemo-dynamical nature of GCs, the BGC metallicity distribution (MD) and comparison of abundances of in situ vs ex situ GCs. 

Here we follow the spirit of the recent investigation by G23, where they derived mean metallicities and radial velocities from medium resolution CaT spectra of a number of members in each of 12 GCs located toward the GB. They then analyzed the nature of their sample and investigated the BGC MDF.

But first we assess the nature of our sample with regards to both in vs. ex situ status as well as classification of in situ GCs as either GB or GD members, in order to ensure the cleanest sample of genuine BGCs as possible. 
In the recent past, we were limited to using relatively weak criteria such as  position on the sky, distance from the Galactic center, and  rather arbitrary metallicity constraints to select what we considered BGCs based on our limited knowledge and understanding of their properties. Bica et al's (2016, 2024 - hereafter B16 and B24, respectively) excellent review articles on BGCs limited their sample to GCs with $R_{GC}<$3kpc and [Fe/H]$\geq -1.5$, with these limits relatively arbtitrarily-imposed (especially the latter).
With the advent of the exquisite proper motions provided by
Gaia, our ability to characterize GCs has been revolutionized
by adding the powerful dimension of kinematics and dynamics. To this, we can add radial velocities to complete our knowledge of the 6-d position-velocity information required to compute orbits for the clusters, derive their integrals of motion and determine their origin.
We note that we will defer to a future paper in this series the reassessment of orbits based on our RVs, and here rely on GC orbital assessments from the literature, since RVs are relatively well determined from previous data.

We have therefore reassessed the nature of potential BGCs quantitatively and have derived a new classification scheme for all Galactic GCs,
both with regards to their in situ or ex situ nature as well as to the bulge or disk status of in situ GCs. We
made use of a number of recent independent sources that we consider to be most relevant. First, following G23, we used Massari et al. (2019 - hereafter M19), Perez--Villegas et al. (2020 - hereafter PV20) and Callingham et al. (2022 - hereafter C22).
M19 used Gaia DR2 kinematical data to derive the integrals of motion of all Galactic GCs, and assigned them to either Main Bulge (MB, 36 GCs), Main Disk (MD) , Low Energy (LE), High Energy, Gaia-Enceladus, Sagittarius, Sequoia, Helmi 99 streams, or XXX (uncertain). We assigned a 1 (in situ) to MB, MD, and LE GCs and a 0 (ex situ) to all other M19 classes. PV20 also used Gaia DR2 kinematics for GCs with $R_{GC}<$4kpc, classifying them as either Bulge/Bar (B/B, 29 clusters), Thick Disk (TD), or Inner/Outer Halo. Again, we assigned a 1 to their B/B and TD GCs and 0 to the rest. C22 used Gaia DR3 dynamics as well as metallicity to assign probabilities to each class based on a multicomponent model, in a more objective and quantitative fashion. They found 42 likely BGCs, and also classified the likelihood of GCs as Disk or Halo, as well as assessed the likelihood of the latter belonging to various accretion events. We simply adopted their membership probability values for each class. 
M19 and PV20 use only dynamical criteria to differentiate in and ex situ objects. We note that
simulations that include
realistic ISM prescriptions suggest that this may be problematic (e.g., Pagnini et al. 2023). Therefore,
we also included results from two more recent studies which employ additional parameters, especially chemistry, in their assessments. Belokurov \& Kravtsov (2023, 2024 - hereafter BK24) use Gaia EDR3 data to derive the integrals of motion of all GCs and calibrate their classification with [Al/Fe] ratios. They find in situ and accreted GCs are well separated 
both chemically and dynamically.
They then simply classify GCs as in situ or ex situ. Finally, Chen \& Gnedin (2024 - hereafter CG24) study clustering  of all GCs in 10 variables, including spatial location, dynamics, metallicity and age, again distinguishing only in  situ and accreted GCs. We simply used the assessments of BK24 and CG24. We then took the mean of these 5 different assessments, giving each equal weight,  and considered a GC as in situ if the mean was $>$0.75.
We limited our initial sample to all MB or LE objects from M19, and added all B/B GCs from PV20 and all with B>0.66 from C22. We found a total of 83 in situ GCs in this sample of 163 GCs, i.e., slightly more than half.

\begin{figure*}
\centering
\includegraphics[width=15cm]{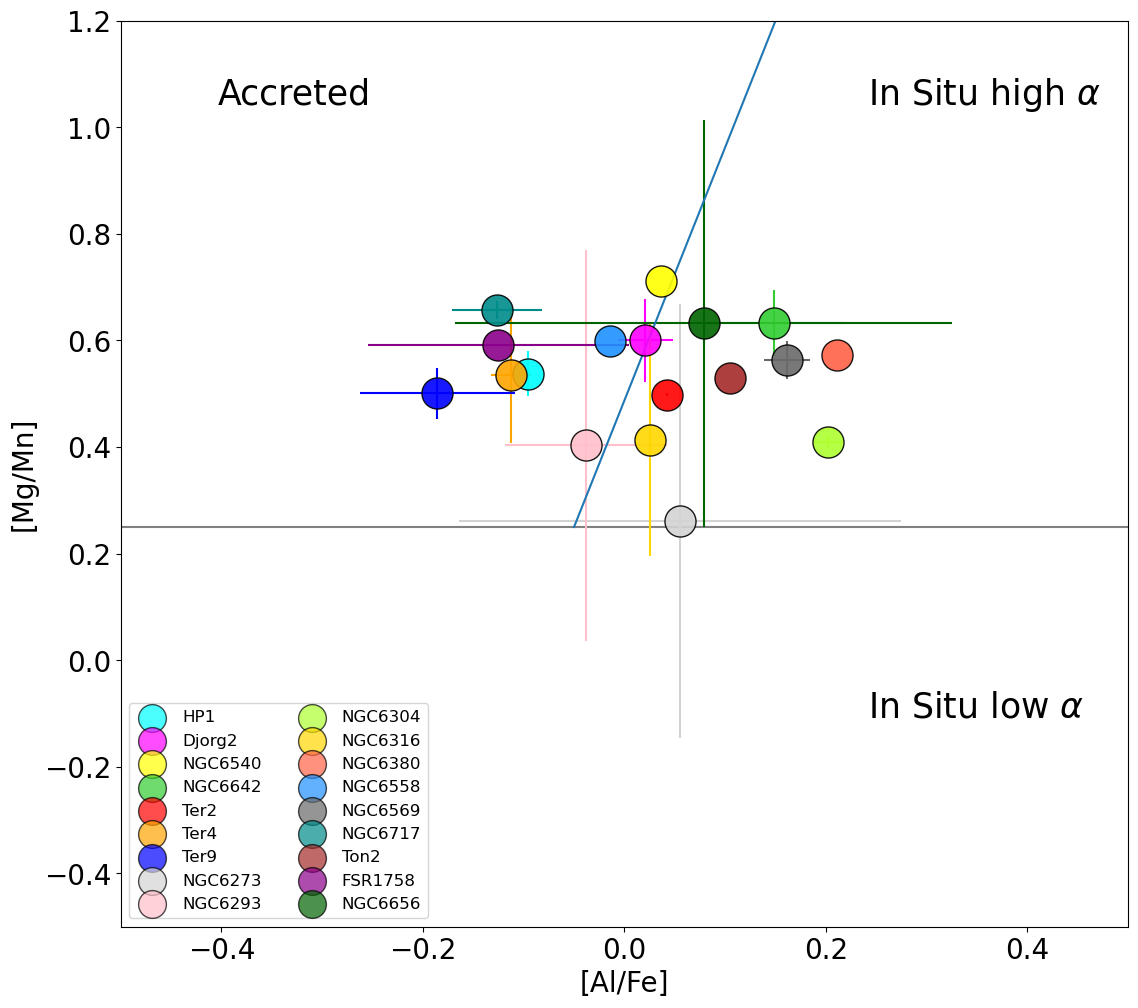}
      \caption{Mean [Mg/Mn] vs. [Al/Fe] for CAPOS clusters. We include the purported division between various origins from Horta et al. (2021).}
    \label{MgMnAlFe}
\end{figure*}

Several additional chemical tests to distinguish in situ vs. ex situ  stars and clusters have been proposed. Horta et al. (2021) uses their location in the [Mg/Mn] vs. [Al/Fe] plane. Accreted stars purportedly inhabit one quadrant, in situ high-$\alpha$ stars another, and in situ low-$\alpha$ yet another. Their Galactic chemical evolution models  showed that there must be some contamination of the “accreted” quadrant by “in
situ” populations.
Vasini et al. (2024) assert that by changing the chemical yields within their error bars, one obtains trends that differ significantly,
making it difficult to draw any reliable conclusion on the star formation history of galaxies from this chemical plane. 
Given this caveat, and the fact that a large fraction of the sample do not have such data available, we prefer not to use this test in our classification. However, for completeness sake, we show this diagram for our CAPOS clusters in Figure \ref {MgMnAlFe}.
We remind the reader that both Mg and Al are elements affected by MP, and that 2P stars are affected by ASPCAP atmospheric parameter issues, so that only 1P stars have been used to define the mean position of our clusters in this diagram, and that the number of such stars is very limited in many clusters. From their position in this plot, HP1, Terzan 4, FSR 1758, Terzan 9 and NGC 6656 all appear as possible/likely accreted GCs, while all other GCs are either in situ high $\alpha$ or on the cusp between this category and accreted stars. 

\begin{figure*}
\centering
\includegraphics[width=15cm]{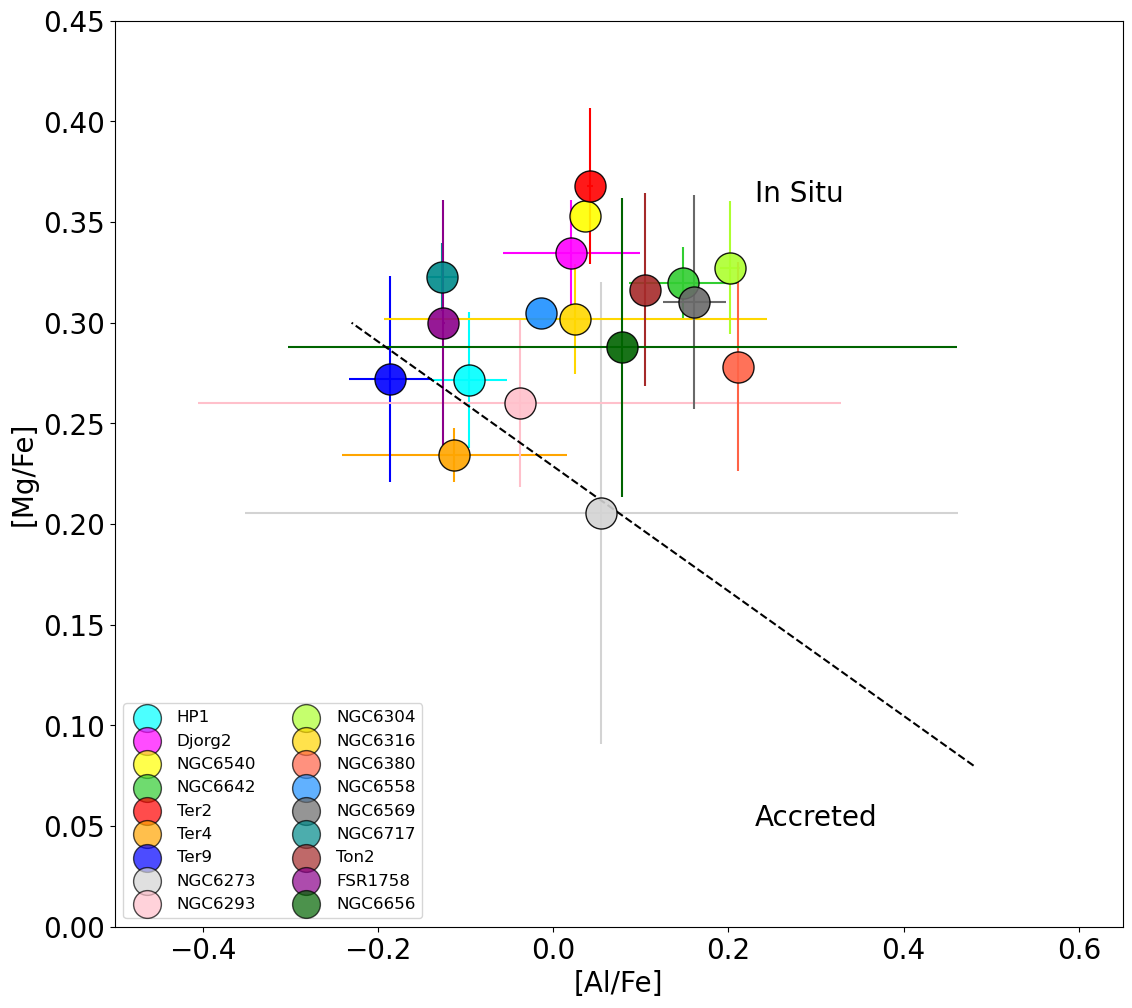}
      \caption{Mean [Mg/Fe] vs. [Al/Fe] for CAPOS clusters. We include the purported division between accreted and in situ clusters from BK24.}
    \label{MgAl}
\end{figure*}

A second  partially chemical distinction has been suggested by  Belokurov \& Kratsov (2023) and BK24. 
 They assume that [Al/Fe] is low in dwarf galaxies due to the metallicity dependence of Al yields and use that information to define the locus of accreted vs in situ populations in integrals of motion space.
We show the mean location of our CAPOS clusters in  their [Mg/Fe] vs. [Al/Fe] space in Figure \ref{MgAl}.  Within the errors, all of the CAPOS sample fall within the in situ region, although Terzan 4, 9 and NGC 6273 are officially slightly within the accreted domain, while FSR 1758, which we classify as our only likely accreted GC, lies slightly within the in situ region.  We again prefer not to use this criterion in our selection process due to the lack of similar data for other clusters and  the ASPCAP problems for Mg and potentially Al, which limit the sample size since we again restrict our sample to 1P stars. In addition, we
note the problems raised by Frelijj et al. (2025) with this test. 

We next similarly quantitatively assessed whether our in situ GCs were either bulge (1) or disk (0)
by assigning a 1 to all MB GCs in M19 and a 0 to all other classes, 
a 1 to all B/B GCs in PV20 and a 0 to the rest, and used the C22 probability value for bulge membership.  We then took the equal-weight mean of these three assessments as our final value, and considered BGCs as those with a B/D mean value $>$0.5. 
Table 6 lists our final BGC sample and their mean B/D ratio.

The number of BGCs we find, 40, compares very well with other recent estimates, e.g., 43 from B16, 
42 from C22
and 42 from B24 (discounting new BGCs), but is based on a number of modern high quality properties and is quantitative, with no prior assumptions regarding metallicity limits, for example. It also uses chemical criteria, at least in some cases, and not exclusively dynamical criteria, to help avoid any issues raised by Pagnini et al. (2023), for example
  that significant overlap between accreted and `kinematically heated' in situ GCs can occur in kinematic space for mergers with mass ratios of 1:10.
However, note that B24 have studied possible new BGCs recently discovered in a variety of surveys such as the VVV, and put together a census of all current likely BGCs, with a total of 61, so our list is incomplete as we do not investigate any of the 39 new BGCs they find, since they have not been subject to the above chemo-dynamical analyses in general.
We find very good agreement with the B24 classification in detail. Only seven of our BGCs are not classified as such by them: two of our GCs fall below their minimum metallicity limit, two are beyond their angular distance limit and a further three lie outside their Galactocentric distance limit.

A second key step to compiling an up-to-date list of BGCs and their salient properties is adding modern, high quality mean metal abundances. A concomitant recent revolution along with Gaia's astrometry has been the collection of high quality metallicities for BGCs, many for the first time, as exemplified by CAPOS. We searched the literature for metallicity determinations of our BGCs, limited to high resolution spectroscopic studies, or CaT measures if no high resolution value was available, and prioritizing the most recent studies. We used APOGEE metallicities (CAPOS  clusters - this paper, Meszaros et al. 2020  for other APOGEE GCs) when available. The results are also given in Table 6, along with the source. 
Note that metallicity spreads are  extensively documented in both Liller 1 and Terzan 5 (Origlia et al. 2013, Taylor et al. 2022, Fanelli et al. 2024), and we have only given the mean value, whereas all other BGCs appear to be monometallic, given the data used.

It is beyond the scope of this paper to compare the variety of metallicity scales quoted here; we simply presume that the scales are in reasonable agreement, as generally expected for modern high quality spectroscopic values. Having a list of such metallicities for BGCs should be a significant improvement over current catalogs, even that of B24, who use an even wider variety of spectroscopic sources and do not prioritize APOGEE.
Unlike B24, we do not attempt to derive $M_V$,
distances or ages for our BGCs, as we feel that in particular the latter two parameters require more homogeneous data than currently available to minimize systematic errors, and refer the reader to B24 for their values.

We carried out the same procedure, including cataloging metallicities, on all other Galactic GCs available from our classification sources, and assembled catalogs of GCs classified as in situ disk (29 GCs), in situ but uncertain as to their disk or bulge nature (14 GCs), and ex situ clusters (61 GCs). We do not include these catalogs here but they are available upon request from the first author.

Only four of the CAPOS sample are not classified as BGCs. NGC 6273 and Ton 2 are considered DGCs, while FSR 1758 and NGC 6656 are found to be ex situ or likely ex situ, respectively. Note that these latter GCs also appear as accreted in Figure \ref{SiFe}. In fact, both of these GCs were added to the CAPOS sample only because they could be observed simultaneously with other GCs that indeed turn out to be bonafide BGCs. 

\begin{table}
\centering
\begin{threeparttable}[b]
\label {BulgeGCs}
\caption{Bulge Globular Clusters}

\centering
\small
\begin{tabular}{ c c c c  }
 
\hline
Cluster ID & B/D ratio  & [Fe/H] &  Source  \\

\hline
NGC 6093 & 0.7 & -1.79 & 9 \\
NGC 6171 & 0.58 & -0.85 & 3 \\
1636-283 (ESO452SC11) & 1 & -0.88 & 13 \\
NGC 6256 & 0.97 & -1.61 & 14 \\
NGC 6266 & 1 & -1.08 & 9 \\
NGC 6293 & 0.65 & -2.12 & 1 \\
NGC 6304 & 1 & -0.49 & 1 \\
NGC 6316 & 0.51 & -0.83 & 1 \\
NGC 6325 & 0.67 & -1.41 & 14 \\
NGC 6342 & 1 & -0.53 & 16 \\
NGC 6355 & 0.66 & -1.39 & 11 \\
Terzan 2 & 1 & -0.88 & 1 \\
Terzan 4 & 1 & -1.41 & 1 \\
HP 1 & 1 & -1.23 & 1 \\
Liller 1 & 1 & -0.21 & 19 \\
NGC 6380 & 1 & -0.90 & 1 \\
Terzan 1 & 1 & -1.26 & 18 \\
NGC 6388 & 0.67 & -0.44 & 3 \\
NGC 6401 & 1 & -1.00 & 2 \\
Terzan 5 & 1 & -0.29 & 19 \\
NGC 6440 & 1 & -0.50 & 5 \\
Terzan 6 & 1 & -0.21 & 2 \\
VVVCL1 & 0.89 & -2.45 & 10 \\
Terzan 9 & 1 & -1.42 & 1 \\
Djorg 2 & 1 & -1.14 & 1 \\
NGC 6522 & 1 & -1.05 & 8 \\
NGC 6528 & 1 & -0.14 & 6 \\
NGC 6540 & 0.99  & -1.09 & 1 \\
NGC 6553 & 0.69 & (-0.12) & 4 \\
NGC 6558 & 1 & -1.15 & 1 \\
NGC 6569 & 0.71 & -1.03 & 1 \\
BH261 & 0.71 & -1.21 & 2 \\
NGC 6624 & 1 & -0.69 & 12 \\
NGC 6626 & 1 & -1.29 & 7 \\
NGC 6638 & 0.66 & -0.99 & 17 \\
NGC 6637 & 0.99 & -0.59 & 17 \\
NGC 6642 & 1 & -1.11 & 1 \\
NGC 6652 & 0.66 & -0.76 & 17 \\
NGC 6717 & 0.99 & -1.17 & 1 \\
NGC 6723 & 0.61 & -0.93 & 15 \\

\hline
\end{tabular}
\textit{1 = this paper, 2 = G23 (CaT), 3 = Meszaros 20, 4 = Munoz 21, 5 = Munoz 17, 6 = Munoz 18, 7 = Villanova 17, 8 = Barbuy 21, 9 = Bailin 19, 10 = Fernandez-Trincado 21b, 11 = Souza 23, 12 = Valenti 11,
13 = Koch 17, 14 = Vasquez 18 (CaT), 15 = Crestani 19, 16 = Johnson 16, 17 =  Carretta 09 (based on Rutledge 97 - CaT), 18 = Valenti 15, 19 = Munoz 25. }

\end{threeparttable}
\end{table}

\subsection{The BGC metallicity distribution}

With our compilation of BGCs and their metallicities in hand, we proceed to investigate their MD. 
Historically, Morgan (1959) and Kinman (1959) showed that the most metal-rich GCs occupy a relatively small volume of (projected) space near the
Galactic center, whereas the metal-poor clusters are spread throughout a much larger volume, the Galactic halo. They first suspected possible distinct GC systems, but no consensus was formed until
Zinn (1985) found that the clusters more metal-rich than [Fe/H] = -0.8 have the properties of a disk system, and first 
established distinct halo and disk systems. Further refinement was performed by
Minniti (1995). On the basis of kinematics, spatial distribution, and metallicity, he argued that the metal-rich GCs within 3 kpc of the Galactic center were more appropriately associated with the Galactic bulge, rather than with the thick disk, thereby establishing the BGC system in addition to the halo and disk systems.

The BGC MD has been recently studied by B16, PV20,  G23, B24 and Garro et al. (2024). As noted above, B16 and B24 defined BGCs as those with $R_{GC}<$3kpc and [Fe/H]$\geq -1.5$, while PV20  used Gaia DR2 astrometry and classified clusters from an orbital analysis. G23 used the classification of Dias et al. (2016).  In addition, B16 used H10 metallicities, while those of PV20 and G23 are mostly from low resolution spectroscopy or even photometry, and B24 from a variety of spectroscopic studies. Thus, both our sample selection criteria as well as their metallicities represent substantial improvements over these prior studies.

\begin{figure*}
\centering
\includegraphics[width=15cm]{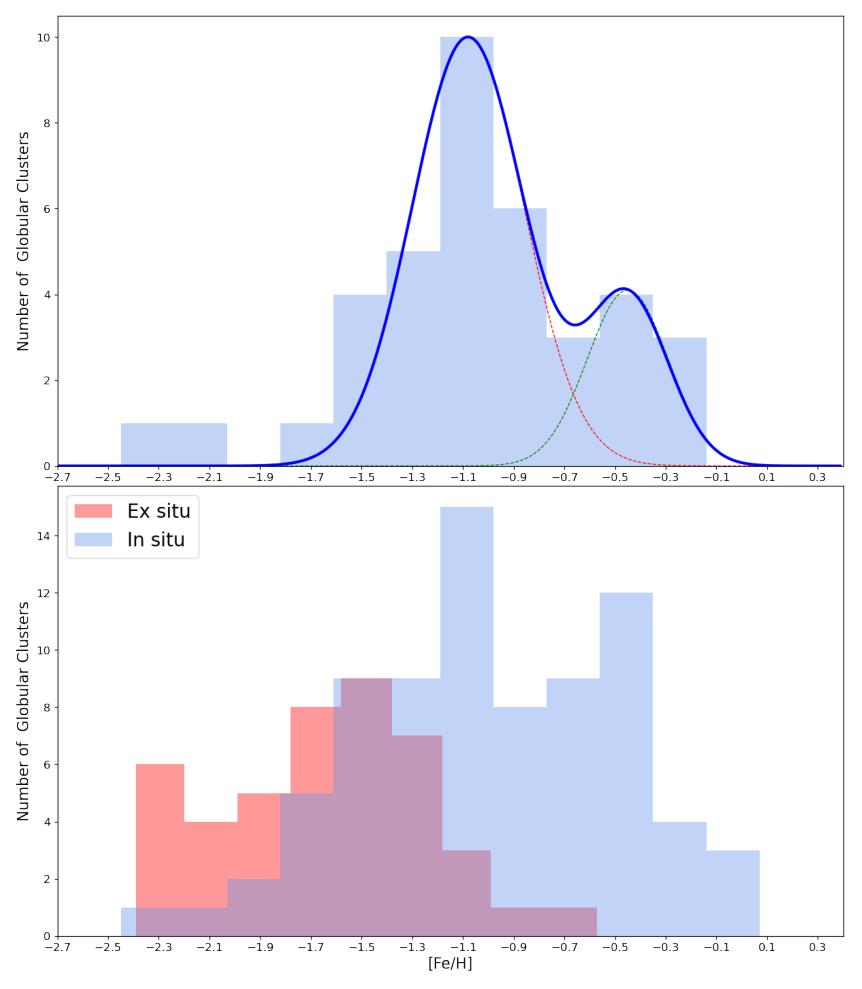}
      \caption{Top: Metallicity distribution (MD) for our BGC sample. The raw histogram and a smoothed fit are shown in blue, while a bimodal gaussian fit is indicated in red and green dashed curves. The BGC MD is strongly bimodal and dominated by the metal-poor peak.
      Bottom: MD for our in situ vs ex situ cluster samples. The in situ sample retains the bimodality displayed by the BGCs alone, while the ex situ MD is mainly unimodal. Both populations have clusters extending to the lowest metallicities, while essentially only the in situ sample has metal-rich clusters.
      }
    \label{MDF}
\end{figure*}

Our MD is shown in Figure \ref{MDF}  (top). Its major features are in fact quite similar to those of the previously mentioned studies. All show bimodality, with one peak around -1.0 to -1.1, and a second peak near -0.3 to -0.5. 
A bimodal gaussian fit to our data yields means of -1.08 with  $\sigma $= 0.24 and 29 GCs in the metal-poor peak, and a mean of -0.45
with $\sigma $= 0.16 and  10 GCs in the metal-rich peak. This compares very well to the two means of -1.05 and -0.4 found by B16, which is the study most similar to ours.

The metal-rich peak is the one that is classically associated with B/D GCs (Zinn 1985).
However, recent MDs generally show there is also a second, more metal-poor peak, which is actually the dominant one. This is especially true of our MD, where we find that more than 70\% of our sample fall in this group. 

A further common feature is the presence of a few BGCs at substantially lower metallicity than the metal-poor peak, extending from $\sim$-1.4 down to -1.8, with two extremely metal-poor GCs below -2.  An important issue raised by G23 was the reality of such low metallicity BGCs. There
is strong interest in determining reliable metallicities and ages for
such metal-poor BGCs, in particular those with a blue horizontal
branch, as they are excellent candidates for the oldest native GCs
of the Milky Way (Lee et al. 1994; Barbuy et al. 2006, 2018;
Dias et al. 2016), since they were born in situ. However, the number of candidates is of course quite small.  We find five clusters with -1.8<[Fe/H]<-1.4: NGC 6325,  NGC 6093, NGC 6256, Terzan 4 and Terzan 9, with only NGC 6093 <-1.61. The first two could potentially be DGCs from their rather low BGC rankings, but the last three clusters appear to be very likely true BGCs.
Finally, our MD exhibits two clusters at even lower metallicities, extending to [Fe/H]$\sim$-2.5. These are NGC 6293 and VVVCL1. From its BGC ranking, NGC 6293 could be a DGC. Figure \ref {MgMnAlFe} also suggests this is reasonable for NGC 6293, the only cluster that is included in CAPOS, but the errors are large. However, the extreme object near -2.5, VVVCL1, ranks as a probable BGC. Neither M19 nor PV20 included this cluster, and B24 labelled it as a halo intruder, simply based on its extreme metallicity.
Both Fernandez-Trincado et al. (2021b) and Olivares Carvajal et al. (2022) perform an orbital analysis based on different parameters, with the former finding that the cluster is probably a  halo GC while the latter prefers it as a BGC. A recent paper (Haro et al.  2025) finds both a higher probability that this is actually a DGC from a new orbital analysis, as well as a somewhat higher metallicity, around -2.25. Vasiliev \& Baumgardt (2021) find it to be a BGC.
In our opinion, these two extreme metallicity outliers are most likely DGCs, since the disk does include other such low metallicity clusters, and the lower metallicity limit of BGCs is $\sim$-1.6, slightly lower than the limit of -1.5 proposed by B16, depending of course on whether NGC 6093 is a bonafide BGC, and, if so, what its metallicity is. This is substantially lower than the minimum  of $\sim$-1.25 suggested by the limited sample of G23. 

At the metal-rich end, there are several BGCs that reach almost solar metallicity, with the two most extreme being  NGC 6528 and NGC 6553, near -0.13 dex. The former in particular is a very probable BGC. A similar metal-rich tail is also seen in the other published MDs. 
Garro et al. (2024) find no super-solar metallicity clusters in the MW, including new clusters in their inner Galaxy sample.
What about the H10 GCs with metallicities >-0.5? There are fourteen that lie towards the bulge (within $20^\circ$) and as such were considered by Zinn (1985) to compose most of his bulge/disk component, as distinguished from the halo, and have come to be 
synonymous with BGCs. Of these, five are classified by our method as non-BGCs, either DGC or indeterminate in vs ex situ objects, while  eight are indeed in our BGC sample (GLIMPSE 2 is unclassified). Thus, the traditional Zinn B/D GCs
that are indeed bonafide BGCs are $<20\%$ of our total BGC population, graphically illustrating how much our classical view of BGCs has changed recently.

A major feature of our BGC MD  is the strong dominance of the -1.1 peak. Most of our sample belongs to this peak, so that in fact a typical BGC has a metallicity between -0.8 and -1.4 and not around -0.5, challenging our paradigm for the canonical metallicity of BGCs.

We caution here that our classification is not definitive and also that, as noted by G23 and proven by B24, the census of BGCs is  still quite incomplete. Near-IR surveys like
VVV/X have uncovered a large number of GC candidates in the
bulge and adjacent disk in recent years (e.g., Minniti et al. 2010;
Moni Bidin et al. 2011; Camargo \& Minniti 2019; Palma et al.
2019; Garro et al. 2022) and some of these, on close inspection,
turn out to indeed be previously
unknown GCs (e.g., Dias et al. 2022). However, some turn out not to be true GCs (e.g., Gran et al. 2019; Minniti et al. 2021; G21; this paper - see Appendix) so further work is required to definitively complete the census of BGCs. B24 indeed find no fewer than 39 new likely BGC members in their recent detailed analysis; these objects demand further investigation.

We finally quantitatively compare our MD to that of bulge field stars. Following G23, we select the recent large APOGEE sample from Rojas-Arriagada et al. (2020). They find strong evidence for trimodality, with peaks at [Fe=H] = +0.32, -0.17,
and -0.66. The fraction
of stars below -1 is very small, in contrast to our sample, where the main peak is more metal-poor than this. The minor BGC peak falls between the two more metal-poor field star peaks.
It is unclear why
the field and GC MDs are so distinct, although an age difference may be a factor. Although GCs are all “old” (i.e.
>10 Gyr) we do not have very accurate ages for the field
stars, which could be somewhat younger, and thus more metal
rich, in general. However, one expects to see the remnants in the bulge field of stars stripped and evaporated from BGCs. We note that bulge RR Lyrae stars peak at around [Fe/H] =
-1.2 (Savino et al. 2020). These stars are of similar old age to the GCs, so it is reassuring that their MDs peak near similar values. However, they do not show a secondary peak near -0.45.  Also, the
MD of 2P field stars identified in the inner Galaxy by Schiavon et al.
(2017b) peaks around [Fe/H]=-1.

\subsection{Metallicity distribution of in situ vs. ex situ GCs}

Armed with our classfication and spectroscopic metallicities of in situ
B/D GCs and their accreted counterparts, it is of interest to compare their MDs. To our knowledge, such a comparison has not been made directly, although G23 plotted separately the MDs of BGCs, DGCs and halo GCs, but based on a different classification scheme.
We note that only 44 of the 61 ex situ sample have metallicities
in our catalog, as we required  high resolution or at least CaT spectra-based metal abundances, while 78 of our 83 in situ GCs have such metallicities, so the ex situ sample is substantially less complete. The comparative lack of good metallicities for accreted GCs compared to in situ GCs is a reversal from what was the case until recently, e.g., in H10. Indeed, CAPOS  has helped significantly in providing such data for previously very poorly-studied objects.

Figure \ref{MDF}  (bottom) compares the in situ vs ex situ MDs. Interestingly, in situ GCs cover the full metallicity range, from solar to almost -2.5, while ex situ GCs are almost exclusively more metal-poor than -1.
The in situ GCs show the same bimodality as the BGCs themselves, not surprisingly since the BGCs dominate, but we note that the DGC MD alone shows the same bimodality. Ex situ GCs are more or less unimodal, with a peak around the classic halo peak of -1.6, but with a substantial number of clusters at the low metallicity end. Although the ex situ GCs dominate below $\sim -1.6$, in situ clusters extend to the lowest metallicity found in the ex situ sample. To first order, metallicity is a decent predictor of the origin of a GC: in situ GCs dominate above $\sim -1.2$, ex situ below $\sim -1.6$,
while in the intermediate regime the probability is more equal.

\subsection{[$\alpha$/Fe] of in situ vs. ex situ GCs}

Following G21, and inspired by Horta et al. (2020), we finally investigate the mean [$\alpha$/Fe] of in situ vs. ex situ GCs. This ratio, as a function of 
[Fe/H], is sensitive to the star formation, and thus chemical enrichment, rate of a system. In a system with a shallower potential well, one expects a lower star formation rate and thus lower [$\alpha$/Fe] at a given [Fe/H], than in a system with a deeper potential. The dwarf galaxy progenitors where the accreted GCs formed presumably
represent a much
shallower potential than the main progenitor where the in situ GCs formed, so one expects the ex situ GCs to have a lower [$\alpha$/Fe] at a given [Fe/H] than their in situ counterparts of similar metallicity. 

This indeed is what G21 and Horta et al. (2020) found. Both studies used Si as a proxy for the  
$\alpha$ abundance to investigate all
GCs well-measured with APOGEE in DR16.
G21 derived a mean [Si/Fe] for their sample
of 11 in situ GCs lying between [Fe/H] = -1 to -1.5 of 0.26$\pm 0.04$
and 0.18$\pm 0.04$ for 13 accreted clusters in the same metallicity range,
where there are a reasonable number of both in situ and accreted
cluster types. The Horta et al. values were almost identical. 

We revisit this issue, again using Si as the 
$\alpha$ proxy, not only for consistency sake but also because of its insensitivity to MP issues, which is not the case for such other possible proxies as Mg, Ca or global $\alpha$.
Note that this paper significantly increases the sample of observed BGCs in CAPOS and also extends them to lower metallicity than the G21 sample.
Besides adding a number of new CAPOS clusters, we can also include a number of other GCs studied with APOGEE in DR17. We employ the Schiavon VAC for these clusters not in CAPOS.  
As a check, we find that the Schiavon [Si/Fe] values are within 0.03 dex  in the mean of  the corresponding CAPOS values for the 18 GCs, as expected since the database is the same.

\begin{figure*}
\centering
   \includegraphics[width=15cm]{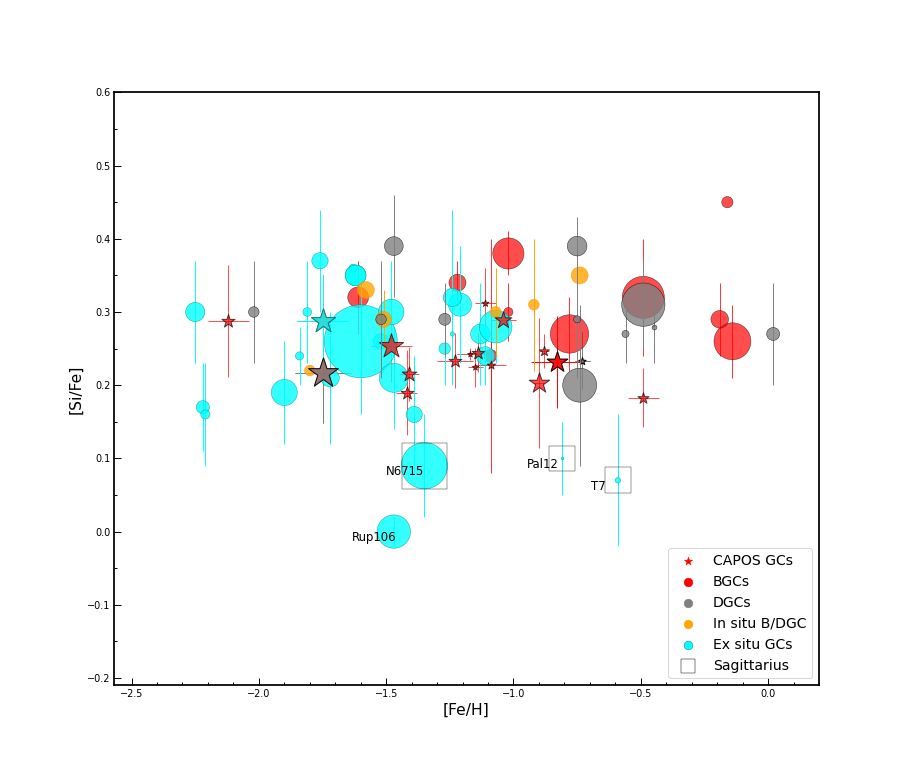}
      \caption{Mean [Si/Fe] versus [Fe/H] for GCs from CAPOS  (red stars), other Bulge GCs (red circles),    other Disk GCs (gray circles), other in situ GCs (unclassified as to whether B or D, yellow circles)
      and ex situ GCs (blue circles).  Values for non-CAPOS clusters are from S24, except for  NGC 6715 (Meszaros et al. 2020), Pal 12 (Cohen et al. 2004), Terzan 7 (Sbordone et al. 2005) and Ruprecht 106 (Villanova et al. 2013).
      Symbol size is proportional to mass.
      We find no significant difference between in and ex situ GCs 
       and no mass dependence.}
    \label{SiFe}
\end{figure*}

Figure \ref{SiFe} shows our results.
Over the entire sample of 44 in situ and  29 ex situ GCs, we find a mean $<[Si/Fe]>_{in situ} =0.28\pm 0.06$ and a mean
$<[Si/Fe]>_{ex situ}$= 0.24$\pm $0.09. For the sample of 17 in situ GCs with Fe abundance between -1 and -1.5, where both Horta et al. 2021 and G21 made their detailed comparison, the mean and $\sigma$ are the same, while the 12 ex situ GCs in this metallicity range also have  virtually the same mean  (0.23) as the full sample  and the same $\sigma $.
Surprisingly, both the full sample as well as the limited  metallicity range sample where the overlap is maximized show  only a small and statistically insignificant difference in the mean Si/Fe abundances.  We note that our sample size is somewhat larger than those of the previous studies.
At first glance, very similar [Si/Fe] abundances at similar [Fe/H] would imply similar SFR/chemical evolution for the in and ex situ samples, in contrast to our expectation that the SFR should be faster in the deeper potential well of in situ vs accreted GCs, leading to higher [Si/Fe] at a given metallicity in the former. Our finding also  is at odds with those of both Horta and G21, who found significant differences between in and ex situ GCs, in  particular in the limited metallicity range. We also find that the GC mass does not correlate with [Si/Fe] and note that our results do not change by using the same ex/in situ classifications as
adopted by G21 and Horta et al. (2020), which are very similar to ours,
 nor do they change significantly if we use the exact same sample as G21 and Horta but now using our results for CAPOS GCs and S24 for other GCs.

If we take our results at face value, one needs to explain why the SFRs of these two distinct GC types should be so similar. One possible explanation is that GCs, which represent the extreme high density tail of star formation, must form in very dense regions, where the local gravitational potential is indeed deep, irrespective of the more global conditions in the galaxy where the GC forms. Also, the merger and star formation histories for dwarfs at high(er) z can be quite diverse, and it may not be the case that GCs from these objects neccessarily have lower 
$[\alpha/Fe]$ at a given [Fe/H] than those formed in the main progenitor.

We note that the GCs with the lowest [Si/Fe] values are all indeed ex situ. We have labeled the 
4 most extremes cases. The most extreme is the unique GC Ruprecht 106, with a solar [Si/Fe]. Ruprecht 106 is possibly unique in another important regard - it may be the most massive GC without MPs (Villanova et al. 2013).  The other 3 extreme cases are all Sagittarius GCs, with a mean [Si/Fe]= 0.09$\pm$0.01, very substantially lower than the other ex situ GCs. Why?

\begin{figure*}
\centering
   \includegraphics[width=15cm]{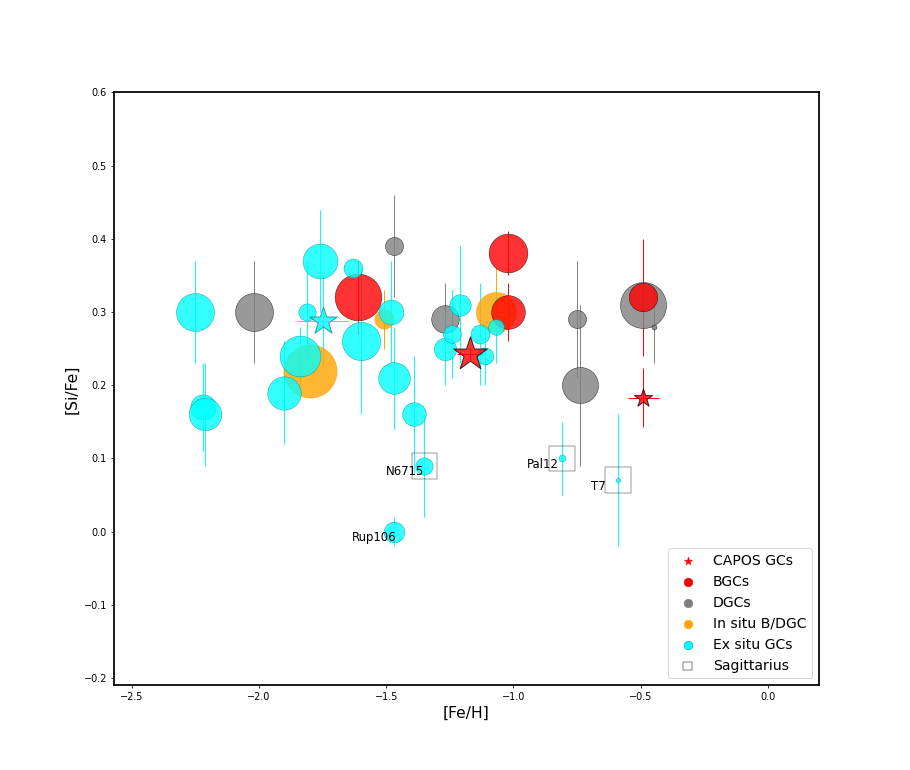}
      \caption{Mean [Si/Fe] versus [Fe/H] for GCs from CAPOS  (red stars), other Bulge GCs (red circles),    other Disk GCs (gray circles), other in situ GCs (unclassified as to whether B or D, yellow circles)
      and ex situ GCs (blue circles).   Values for non-CAPOS clusters are from S24, except for  NGC 6715 (Meszaros et al. 2020), Pal 12 (Cohen et al. 2004), Terzan 7 (Sbordone et al. 2005) and Ruprecht 106 (Villanova et al. 2013).
      Symbol size is proportional to age from Valcin et al. (2025). Other than the 4 most extreme low [Si/Fe] GCs, which are relatively young and generally members of Sagittarius, no strong age dependence is observed.} 
      
    \label{SiFeAge}
\end{figure*}

 A possible explanation could be age.
Theoretically, similar high [$\alpha$/Fe] abundances would be expected in all GCs despite their origin IF they were all born long ago, within a Gyr or so of the initial burst of relatively primordial star formation in their respective protogalaxy, after initial SNII had exploded but before any SNIa had gone off to decrease [$\alpha$/Fe].
It is well known that most GCs are indeed very old,
of the order of 13Gyr or more (e.g. Valcin et al. 2025). However, at least some of them are known to be somewhat younger, and one would expect that GCs a Gyr or more younger than the rest might be expected to show lower [$\alpha$/Fe]
from subsequent SNIa ejecta into the ISM before their formation. Is there any discernible age difference between high and low [$\alpha$/Fe] GCs? We use the recent large-scale homogeneous age compilation of Valcin et al. (2025) to address this,
shown in Fig \ref{SiFeAge}. This is identical to Fig \ref{SiFe}
except the symbol size is now directly proportional to age instead of mass. 
Note that the number of CAPOS GCs is greatly decreased because their ages are poorly determined and not in the Valcin paper. There is no overall clear separation of [Si/Fe] with age - the 19 GCs older than 12 Gyr have a mean [Si/Fe]=0.27$\pm$0.06 while the 20 younger GCs have a mean [Si/Fe]=0.23$\pm$0.10, which is again only a small and statistically insignificant difference.  
However, the 4 extreme cases noted above are all quite young, which could explain their low [$\alpha$/Fe].

A closer look reveals some potential analysis issues. Horta et al. used APOGEE ASPCAP data in DR16, while S24 used ASPCAP data in DR17. How do these two analyses compare? We find that the Horta et al.  mean [Si/Fe] values for 20 GCs in the limited metallicity range of interest are on average 0.08 dex lower than the corresponding S24 values. This might explain part of the discrepancy. However, this begs the question:
why are the Horta and S24 values for the same clusters offset? Are the DR16 and DR17 values so different? 
A  possible partial cause could be the S24 membership   strategy that we confirmed does lead to field star contamination in some cases. However, we only found these in some of our CAPOS sample, and have not examined other GCs. 
A further complication is that we found BACCHUS [Si/Fe] abundances to be significantly higher than ASPCAP values for our CAPOS clusters with both analyses in hand. Clearly, more careful work is needed to help solve these puzzles and definitively compare in and ex situ GCs in this important diagnostic.

\section{Conclusions}
We present final ASPCAP results for all 18 GCs observed in the CAPOS External Program as part of SDSS-IV, building on the initial results given by G21 for a subsample of GCs and member stars. CAPOS was designed to obtain detailed abundances and kinematics for as complete a sample of bonafide BGCs as possible using the unique advantages of APOGEE in order to exploit their extraordinary Galactic archaeology attributes.
We utilize ASPCAP data from DR17 to investigate atmospheric parameters, detailed abundances and velocities for the sample.

We first carry out a stringent systematic selection of members, using the criteria of position within the tidal radius, proper motion, metallicity, radial velocity and CMD location to ensure very high membership probability for our final surviving targets. We find a total of  303 members with SNR$>$70, for a mean of  almost 17 high-quality stars per cluster,
 and an additional 125 stars with lower SNR.
In this process, we found good general agreement but some discrepancies with the members selected using a similar procedure on DR17 data by Schiavon et al. (2024), in the sense that they included a number of stars that were rejected by our procedure in several clusters.  Such discrepancies are to be expected given their intentionally generous membership criteria.

Following G21, we then investigated the reliability of the ASPCAP atmospheric parameters but with our much larger sample. We reinforced and clarified the finding of G21 that stars with the highest [N/Fe]
abundances within a given cluster show a higher [Fe/H] than their cluster
counterparts with lower [N/Fe]. The
1P stars behave as expected but 2P stars exhibit a roughly linear increase in [Fe/H] with [N/Fe], with the limit reasonably well defined at [N/Fe] = +0.7. Only our most metal-rich cluster does not show any trend. 
We corrected the [Fe/H] value of 2P stars in a cluster using this trend and combined them with the uncorrected [Fe/H] values of 1P stars to derive the final cluster mean metallicity. 

We also examined the behaviour of other key elements in this regard, finding that [Mg/Fe], [Ca/Fe] and $[\alpha /Fe]$ all exhibit significant trends with [N/Fe] for 2P stars, but that [Si/Fe] shows only a very small systematic deviation for even
the most extreme 2P stars, while 1P stars are unaffected in all these elements. We conclude that we can safely use the ASPCAP [Si/Fe] values for all stars but will limit our Mg, Ca and  $\alpha$ abundances, as well as abundances for all other elements except Si, to 1P stars only,  which should all be well determined by ASPCAP. We also decided to use Si as the best surrogate for the overall $\alpha$ abundance.

We then proceeded to derive mean abundances for all elements in each cluster, as well as the mean radial velocity, and compared our values to the most relevant literature results. 
Our mean metallicities are determined to within 0.05 dex (mean standard deviation), our mean $[\alpha /Fe]$ (= [Si/Fe]) to 0.06 dex, and our mean radial velocity to 3.4km/s.
None of our clusters show strong evidence for internal metallicity variations, although the sample size in some clusters is relatively small, limiting the statistics. This is not the case for NGC 6656, which is the subject of a longstanding debate in the literature about a possible metallicity spread. Our sample of 130 high SNR members in NGC 6656 shows a standard deviation in metallicity of only 0.10 dex, mimicking the finding of Meszaros et al (2020) based on a much smaller sample using APOGEE and BACCHUS, who also did not support an intrinsic metallicity spread. We note that a future paper in this series will be devoted to a BACCHUS analysis of the full sample of some 250 NGC 6656 APOGEE red giant members with SNR$>$70 which should definitively address this issue. 

Our mean metallicities are generally in good to
very good agreement with previous high quality spectroscopic
studies but these latter are limited due to the high extinction that most of our sample suffers. We emphasize that the near-IR, high resolution
and high SNR observations, relatively large sample sizes and excellent
membership probabilities, and homogeneous abundance
derivations make our CAPOS results unparalleled.

We find that our mean 
 [Fe/H] values are within 0.05 dex but
[Si/Fe] values are  0.12 dex lower than those derived for the same  11  CAPOS clusters that have been studied to date using BACCHUS and independent photometric parameters. The nature and cause of the  [Si/Fe] offset will be further investigated in future papers in this series studying CAPOS clusters with BACCHUS.

Our mean radial velocities are generally in very good to excellent agreement with B19 except for Terzan 9 and NGC 6642, where we agree with G23 that the B19 values for these two GCs are highly suspect. A check on the new DR3 values for these two GCs shows that the discrepancy is now gone.
We find good general agreement between our bulge
cluster and field star samples in both Mn and Ni. 

To best explore the most salient implications of CAPOS, we first assess the nature of our sample with regards to both in vs. ex situ status as well as classification of in situ GCs
as either GB or GD members,  to ensure the cleanest
sample of genuine BGCs as possible. We derived a new chemo-dynamical classification scheme for
all Galactic GCs, synthesizing the results of a number of recent classification studies.
Slightly more than half of the 163 GCs we studied are classified as in situ. Forty of these are deemed BGCs, which matches very well with other recent assessments. However, we note that our list is undoubtedly incomplete 
as we do not investigate any of the 39 new BGC clusters/candidates recently compiled by B24.
All but four of our CAPOS sample are classified as BGCs.
We also compile an up-to-date list of modern, high quality mean
metal abundances, using APOGEE values when available, or other high resolution spectroscopic studies or finally CaT measures if no high resolution value was available.

We first investigate the MD of our BGCs. As in other similar recent studies, the BGC MD is strongly bimodal, with peaks near -1.1 and -0.45. The latter is the one traditionally associated with BGCs; however, not only is there a second, more metal-poor peak but it is in fact the dominant one, containing  $>70\%$ of all BGCs.  A short tail extends from the higher metallicity peak up to near solar, and a longer tail extends to much lower metallicity than the metal-poor peak. We suggest that the lower metallicity limit is near -1.6, and that the two "BGCs" we find at much lower metallicity, NGC 6293 and VVVCL001, are most likely DGCs. Improved distances and orbits for these clusters is required to determine their true nature. The latter object is a particularly intriguing GC, with a mean metallicity $\sim$ -2.25
(Haro et al. 2025), thus likely the most metal-poor, and thus likely also the oldest, in situ GC in the MW, with even the possibility of a metallicity spread (Haro et al. 2025). More high resolution near-IR spectra are required to pin down its mean metallicity, potential spread and radial velocity, as well as deep near-IR photometry to derive the age for this possibly oldest main progenitor GC. Our group is working on all of these key goals.

We next compare the MD of in vs ex situ GCs. The in situ GCs
show the same bimodality as the BGCs themselves, while the ex situ GCs are more or less
unimodal, with a peak around the classic halo peak of -1.6, but
with a substantial number of clusters at the low metallicity end. To first order, metallicity is a decent predictor of the origin of a GC: in situ GCs dominate above $\sim -1.2$, ex situ below $\sim -1.6$,
while in the intermediate regime the probability is roughly equal.

Finally, we investigate the mean $[\alpha /Fe]$ of in situ vs. ex situ GCs. Theoretically, one expects the ex situ GCs to have a lower $[\alpha /Fe]$ at a given [Fe/H] than their in situ
counterparts of similar metallicity, given their likely origin in dwarf galaxies with shallower potential wells and thus lower star formation rates compared to the main progenitor,
which has been confirmed by several previous studies, including G21. Here, as in the previous studies, we use [Si/Fe] as the best proxy for $[\alpha /Fe]$. Note that this paper significantly increases the
sample of observed BGCs in CAPOS, now using all of the CAPOS observations and DR17, and also extends them to
lower metallicity than the G21 sample. Surprisingly, we find   only a small and statistically insignificant difference in the mean [Si/Fe] abundances of in and ex situ GCs. We point out possible theoretical explanations for this  e.g. younger ages, at least for the lowest [Si/Fe] clusters, which are also ex situ and mostly in Sagittarius, but find no strong correlation and also find some inconsistencies in various analyses of  [Si/Fe] abundances, and urge further careful analysis with as homogeneous a sample as possible to clarify this conundrum.

In an Appendix, we show that all three of the Minni candidate GCs observed by CAPOS show no clustering in either proper motion, CMD or metallicity:radial velocity space and are therefore not star clusters of any kind.

For the near future, we first note that all of our CAPOS clusters are being analyzed with the BACCHUS package using independent photometric atmospheric parameters to mitigate ASPCAP issues with 2P stars, which are prevalent in our sample. We also
note that there is an ongoing SDSS-V Open Fiber project to observe 
all the remaining BGCs not
observed by APOGEE in SDSS-III or SDSS-IV in order to  extend CAPOS and obtain a
complete sample.

\begin{acknowledgements}
We would like to warmly acknowledge SDSS-IV and especially all of those who have contributed to the outstanding success of both APOGEE North and South.  The referee, Ricardo Schiavon, provided a very detailed and in-depth report which greatly improved this paper.
D.G. and S.V. gratefully acknowledge the support provided by Fondecyt regular no. 1220264.
D.G. also acknowledges financial support from the Dirección de Investigación y Desarrollo de
la Universidad de La Serena through the Programa de Incentivo a la Investigación de
Académicos (PIA-DIDULS), as well as the moral and amorous support of M.E. Barraza. 
C.M. thanks the support provided by  ANID-GEMINI  Postdoctorado No.32230017.
D.M. acknowledges support from the Center for Astrophysics and Associated Technologies CATA by the ANID
BASAL projects ACE210002 and FB210003, and by Fondecyt Project No. 1220724. 
AM acknowledges support from the FONDECYT Regular grant 1212046, from the ANID BASAL project FB210003, as well as funding from the HORIZON-MSCA-2021-SE-01 Research and Innovation Programme under the Marie Sklodowska-Curie grant agreement number 101086388.
C. Montecinos is supported by the National Agency for Research and Development (ANID)  through Programa Nacional de Becas de Doctorado (DOCTORADO BECAS CHILE/2022 - 21220138).
AMK acknowledges support from grant AST-2009836 and AST02408324 from the National Science Foundation.

\\\\

This work has made use of data from the European Space Agency (ESA) mission {\it Gaia} (\url{https://www.cosmos.esa.int/gaia}), processed by the {\it Gaia} Data Processing and Analysis Consortium (DPAC, \url{https://www.cosmos.esa.int/web/gaia/dpac/consortium}). Funding for the DPAC has been provided by national institutions, in particular the institutions participating in the {\it Gaia} Multilateral Agreement.
\\\\
Based on observations obtained through the Chilean National Telescope Allocation Committee through programs  CN2017B-37, CN2018A-20, CN2018B-46, CN2019A-98 and CN2019B-31.
\end{acknowledgements}

%
%




\appendix
{\bf Appendix. Investigating the reality of Minni candidate globular clusters}

As noted in G21, CAPOS also targeted any recently discovered BGC candidates that could be observed serendipitously simultaneously with cataloged BGCs, given the large field of view of APOGEE, as supplemental targets after all fibers that could be placed on known BGC stars have been assigned. We present here our investigation of such candidates, finding that none of them are bonafide clusters.

Appropriate targets satisfying the above were taken from the list of candidate GCs found in the VVV survey by Minniti et al (2017b) and Palma et al. (2019). These are referred to as Minni candidates. We were able to include one candidate in each of three CAPOS fields: Minni 6 in field 000-06, Minni 51 in field 003-03 and Minni 66 in field  350-03. The number of spectra obtained varied due to the nature of the field, how many other known GCs were included, etc. In the end, we were able to obtain good spectra for 46 stars in the field of Minni 06 with SNR between 75 to 360, 15 in Minni 51 with SNR between 8 to 378  but only 4 in Minni 66 with SNR between 64 to 194.

To assess the possible cluster nature of the candidates, we selected stars using the coordinates and radius listed in Minniti et al. (2019, 2017c, 2017d) and compared the selected stars using these criteria with different values of proper motion, RV, metallicity, and color, as well as with field stars.

Figures 1-3 show our results. We find no clustering in proper motion, CMD or metallicity: RV space for any of these candidates and thus no evidence for their reality as clusters. Note that G21 investigated a smaller sample in Minni 51 and arrived at the same conclusion.

\begin{figure*}
\centering
\includegraphics[width=11cm]{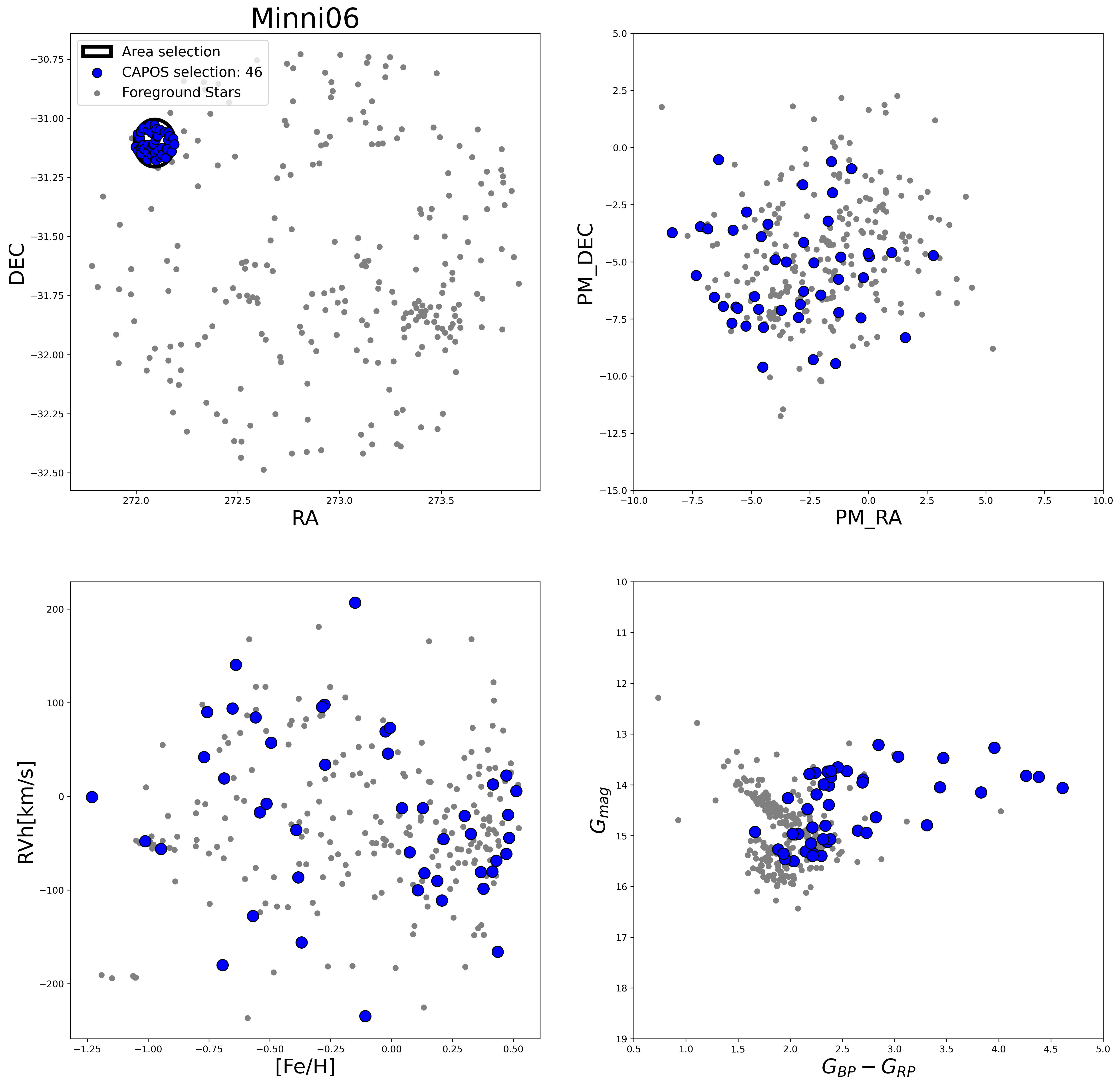}
      \caption{Analysis of the possibility that Minni 6 is a real cluster. Parameters used are the same as in  Figure \ref{N6380members}. No clustering is found in either PM, RV, metallicity or CMD space.}
    \label{Minni6}
\end{figure*}

\begin{figure*}
\centering
\includegraphics[width=11cm]{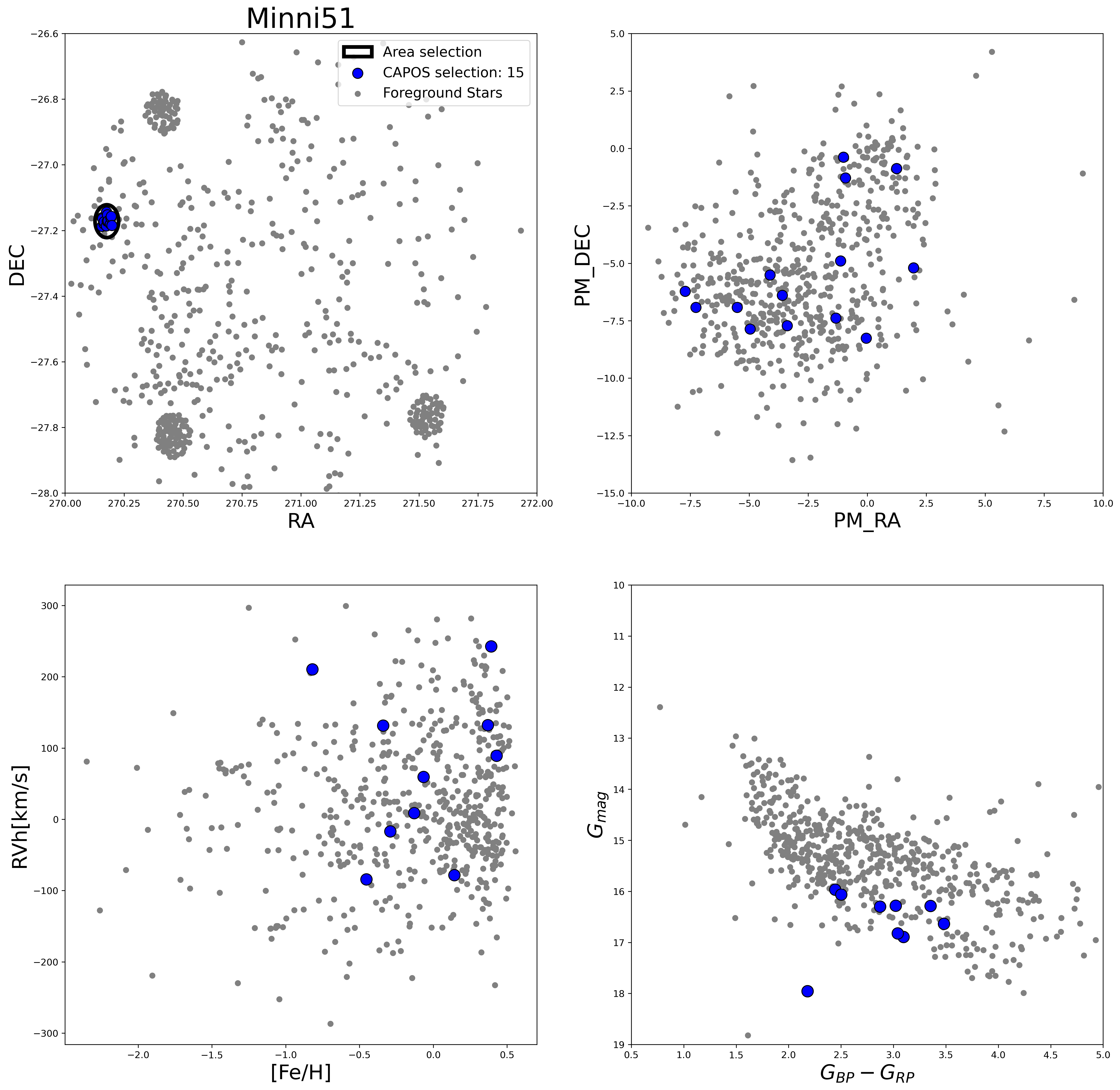}
      \caption{As above for Minni 51.}
    \label{Minni51}
\end{figure*}

\begin{figure*}
\centering
\includegraphics[width=11cm]{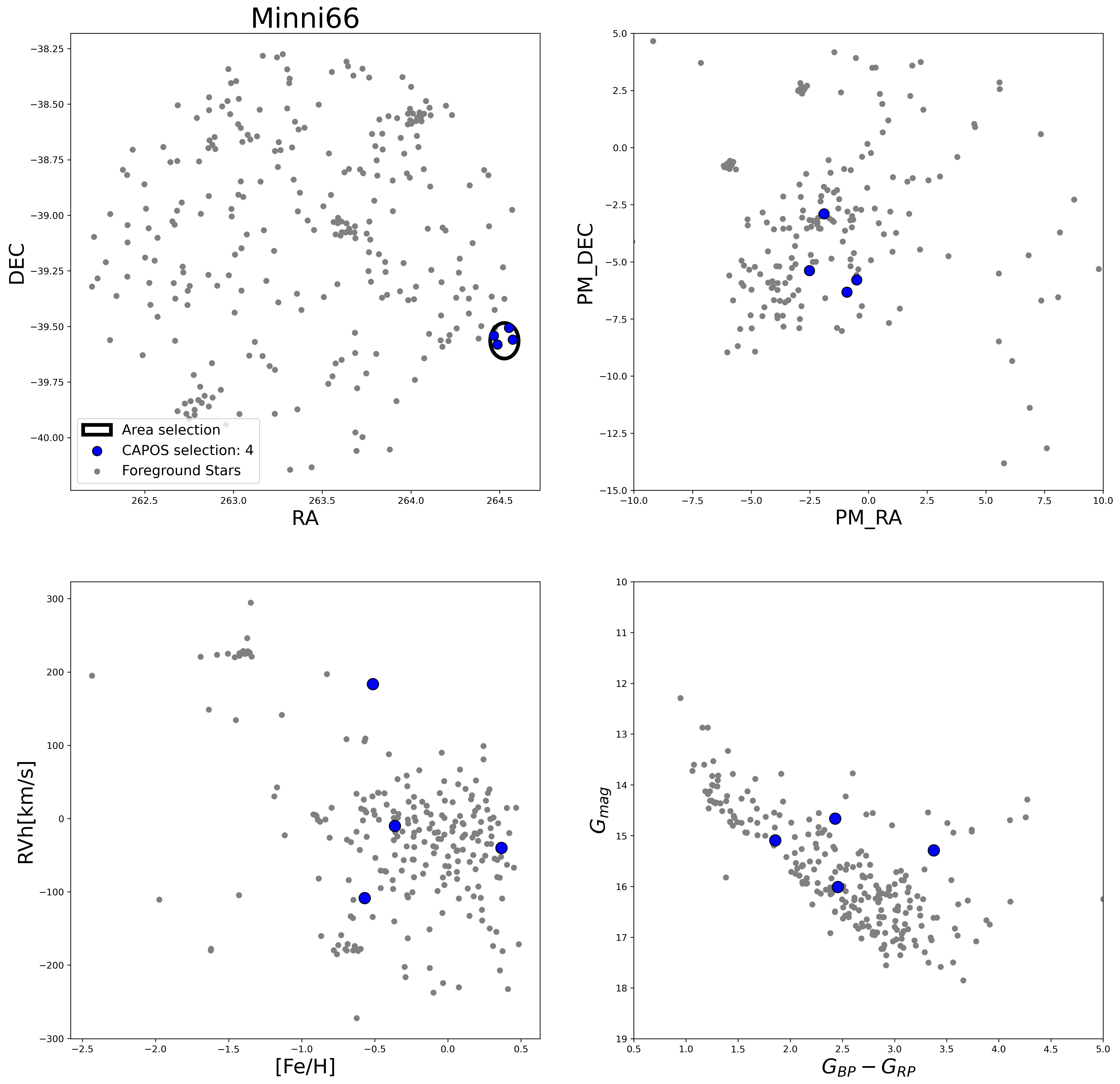}
      \caption{As above for Minni 66.}
    \label{Minni66}
\end{figure*}

\end{document}